\documentclass[journal]{IEEEtran}

\usepackage{xcolor,soul,framed} 

\colorlet{shadecolor}{yellow}
\usepackage{graphicx}
\usepackage{epstopdf}
\graphicspath{{../pdf/}{../jpeg/}}
\DeclareGraphicsExtensions{.pdf,.jpeg,.png}

\usepackage[cmex10]{amsmath}
\usepackage{array}
\usepackage{mdwmath}
\usepackage{mdwtab}
\usepackage{eqparbox}
\usepackage{url}
\usepackage{subfigure}
\usepackage{amssymb}
\usepackage{amsthm}
\usepackage{cite}
\theoremstyle{definition}

\newtheorem{corollary}{Corollary}

\newtheorem{lemma}{Lemma}
\newenvironment{myproof}{{\noindent\it Proof:}}{\hfill $\blacksquare$\par}

\begin{document}
\bstctlcite{IEEEexample:BSTcontrol}
    \title{Beyond Near-Field: Far-Field Location Division Multiple Access in Downlink MIMO Systems}
  \author{Haoyan~Liu, 
      Jianjie~Cai,
      Min~Yang,
      and~Chengguang~Li
      \thanks{The authors are with the School of Aerospace Science and Technology, Xidian University, Xi’an 710071, China(e-mail:liuhy@xidian.edu.cn; 22131214175@stu.xidian.edu.cn; merovingia1911@126.com; chgli@stu.xidian.edu.cn)}}  

\maketitle

\begin{abstract}
Space division multiple access (SDMA) with massive multiple-input
multiple-output (MIMO) has been crucial in enhancing system capacity through spatial differentiation of users. However, SDMA can only increase spatial resolution in ultra-dense networks through the deployment of extremely large-scale antenna arrays. For a long time, research has predominantly focused on the angle dimension, overlooking the potential of the distance dimension. Benefiting from the beam-focusing effect of the near-field spherical propagation model, near-field location division multiple access (NLDMA) was proposed to partition channel resources across both angle and distance domains. To extend this concept to the far-field region, this paper introduces a far-field LDMA (FLDMA) scheme utilizing frequency diverse arrays (FDA), which endow the arrays with distance resolution capabilities. Specifically, we optimized the orthogonal frequency division multiplexing (OFDM) waveforms to ensure that the subcarriers maintain orthogonality under FDA reception conditions. Additionally, the designed steering vectors satisfy asymptotic orthogonality in the two-dimensional plane and average out the inter-user interferences. Theoretical and simulation results demonstrate that FLDMA can fully exploit the additional degrees of freedom in the distance domain to significantly improve spectral efficiency, especially in narrow sector multiple access (MA) scenarios. Moreover, the independence of array elements can be maintained even in single-path channels, making FLDMA a standout choice among MA schemes in millimeter-wave and higher frequency bands.

\end{abstract}


%
\IEEEpeerreviewmaketitle


\section{Introduction}
\subsection{Background and Motivation}
The three primary application scenarios defined in 5G—enhanced mobile broadband (eMBB), massive machine-type communications (mMTC), and ultra-reliable low-latency communications (URLLC)—will encounter increasing challenges in the next generation of wireless communication networks. In 6G, eMBB is expected to support more demanding applications, such as immersive holographic communications and ultra-high-definition video streaming, necessitating significant advancements in spectral efficiency. Simultaneously, mMTC will expand to accommodate the explosive growth of Internet of Things (IoT) devices in smart living and industrial production. Moreover, URLLC will provide real-time control and coordination capabilities for critical applications requiring stringent latency and reliability, such as autonomous vehicle networks and the full automation of interconnected robots in ultra-dense networks \cite{8869705}. Multiple access (MA) schemes play a crucial role in enhancing service quality in multi-user scenarios. Therefore, it is imperative for novel MA technologies to continuously evolve to meet the demands of 6G scenarios, further enhancing system capacity, increasing connection density, reducing power consumption and costs, and minimizing access delays \cite{clerckx2024multiple,liu2024road}.

In previous generations of mobile communication systems, MA schemes such as time/frequency division multiple access, code division multiple access, and orthogonal frequency division multiple access have been employed to separate users by dividing orthogonal channel resources. The evolution to 4G and 5G systems witnessed the emergence of space division multiple access (SDMA), leveraging massive multiple input multiple output (MIMO) technology \cite{6736761,7414384}. The base station (BS) employs advanced precoding methods to allocate each user equipment (UE) a unique steering vector or beam, thereby enabling angle division of users \cite{1415913,1683918,4712693}. Despite the residual inter-user interferences (IUIs) after precoding, the multiplexing of time-frequency resources has significantly increased system capacity. Furthermore, power domain non-orthogonal multiple access has garnered substantial attention in 5G, demonstrating the potential of transmitting superimposed messages from multiple UEs within the same channel resource block \cite{6692652}. It is evident that the expansion of channel dimensions has markedly accelerated the evolution of communication systems. Contemporaneous schemes, including rate-splitting multiple access \cite{9831440}, pattern division multiple access \cite{7526461}, and sparse code multiple access \cite{6666156}, have optimized interference management strategies from different information processing perspectives, leading to more effective utilization of existing channel resources. However, these methods lack further development of the available channel dimensions. Some novel MA strategies for 6G have begun exploring the resource dimensions at the backend of communication systems. For instance, Pradhan et al. have applied finite field extensions to MA \cite{9797778}, and Zhang et al. have proposed mode division multiple access based on semantic domain resources \cite{2023Model}.

It is apparent that there is yet another dimension in the space that has rarely been considered: distance. To our best knowledge, Wu and Dai were the first to develop the channel degrees of freedom (DoFs) in the distance domain and proposed the innovative near-field location division multiple access (NLDMA) \cite{10123941,10243590}. For extremely large-scale antenna arrays, the Rayleigh distance extends to several tens of meters, indicating that a significant portion of communications will occur in the near-field region characterized by spherical wave propagation. By utilizing the near-field propagation model, beams can focus signal energy within specific locations of a two-dimensional plane \cite{10.1063/1.5011063}. This capability allows the BS to divide UEs effectively based on their angles and distances. The beam's focusing capability progressively diminishes as the distance increases, ultimately degrading into planar waves. Thus, the implementation of NLDMA is extremely dependent on large-scale arrays and high-frequency electromagnetic waves to extend the Rayleigh distance. Nonetheless, NLDMA has demonstrated significant improvements in spectral efficiency through the exploitation of the distance dimension. Likewise, a new challenge arises: how to focus beams in the far-field?

\subsection{Related Work on Frequency Diverse Array (FDA)}
The foundation of radar ranging lies in utilizing signal propagation delay, an aspect often overlooked in communication systems. To endow arrays with distance perception capabilities, FDA introduces slight frequency offsets to each array element \cite{1631800}. Consequently, waves from different elements yield phase differences during propagation, leading to interference patterns at specific distances upon superposition. The original FDA faces two major challenges. The first challenge stems from linear frequency increased offsets, which produce oscillating S-shaped beampattern in the two-dimensional plane. This limitation can be mitigated by employing logarithmic increased \cite{8649595}, random \cite{7740037}, or optimization-based \cite{9219125} frequency offsets, thus decoupling angle and distance dimensions to form a dot-shaped beampattern. The second key issue is the time-variant beampattern of the FDA, preventing the peak from consistently orienting toward the target location. Many efforts have attempted to overcome this obstacle by designing time-modulated frequency offsets \cite{7522082} or transmit weights \cite{7877873}. However, these approaches are fundamentally flawed as elucidated in \cite{10145045}. The time-variance is an inherent characteristic resulting from the various frequency offsets in FDA, making it impossible to generate static beams in the distance-angle domain as achieved by phased array (PA). A feasible approach is to utilize the joint transmit-receive schemes to generate an equivalent time-invariant beampattern, by setting a group of matched filters in each channel of the receiver \cite{8550659}. In digital communication, the primary concern is the baseband model, rather than ensuring that beams maintain directionality toward the UE. Consequently, the joint transmit-receive FDA smooths the way toward the utilization of the distance domain.

Benefiting from the brand new dimension provided by the FDA, many studies have begun exploring its applications in communication systems. \cite{8765358,10171376,jian2023fdamimobased} proposed integrated sensing and communication schemes based on FDA through embedding information into frequency increments to minimize the impact of communication functions on sensing performance. Subsequently, some studies have incorporated index modulation into FDA-MIMO systems to enhance communication rates \cite{10272262,9364868}. The most pertinent study by Jian et al. explores FDA-based beamforming to transmit secrecy information from the BS to downlink UEs \cite{10064127}. Since activating the range-dependent property of FDA requires considering propagation delay at the receiver \cite{9171574}, it will inevitably introduce symbol timing offset (STO) and consequently reduce the orthogonality of the matched filter during symbol time integration. In narrowband communication systems, the propagation delay is much shorter than the symbol time, as the aforementioned studies have neglected the issue of orthogonality degradation. However, it will span several tens of symbols in wideband systems. Therefore, 
it is essential to employ waveforms that can accommodate propagation delay to support the FDA in high-rate communication scenarios.

\subsection{Our Contributions}
In this paper, we design a far-field LDMA multiple access (FLDMA) scheme for the downlink by exploiting the distance resolution capability of FDA in wideband systems. Compared to SDMA, the proposed scheme can effectively mitigate interference between UEs in nearby directions by fully exploring the additional DoFs in the distance domain, thereby improving spatial performance. However, the frequency offsets introduce both additional spectral overhead and inter-carrier interferences (ICIs) at the receiver. Hence, whether FLDMA can achieve an overall improvement in spectral efficiency compared to SDMA remains to be further investigated. The main contributions can be summarized as follows:
\begin{itemize}
    \item  Inspired by the FDA technique, the development of FLDMA is explored for wideband systems. The key idea is that the inherent cyclic prefix (CP) of orthogonal frequency division multiplexing (OFDM) allows the receiver to maintain subcarrier orthogonality even in the presence of STO. Moreover, the matched filtering module of the OFDM receiver can simultaneously eliminate the time-variance of the steering vectors and demodulate the information symbols carried by subcarriers. These advantages ensure that the proposed MA scheme does not require any modifications to the receivers on the UE side.
    \item The proposed random permutation frequency offsets effectively decouple the distance-angle dependency while exhibiting discrete Fourier transform properties in the distance domain, similar to those in the angle domain. We further demonstrated the asymptotic orthogonality of the designed steering vectors. Subsequently, two methods for eliminating ICI are examined. While small frequency offsets can approximately maintain subcarrier orthogonality, this approach leads to a beamwidth that exceeds the maximum propagation distance supported by the designed waveform. In contrast, although the pre-equalization method introduces additional complexity, it enables the design of narrower beams in the distance domain, which enhances the BS's capability to resolve UEs at different locations more effectively.
    \item The upper bound of FLDMA performance has been demonstrated. When the number of UEs is small, such as only two, its performance typically does not surpass that of SDMA. However, as the number of UEs increases, FLDMA achieves significant improvements in spectral efficiency by trading off a portion of spectral resources. Simulation results indicate that FLDMA can achieve up to a 75\% performance improvement when the sector angle is large; for small sector angles, FLDMA's spectral efficiency is several times that of SDMA. Mathematical derivation further elucidates the reasons behind the substantial gains of FLDMA: traditional SDMA requires a rich scattering environment to ensure strong independence between array elements \cite{Tse_Viswanath_2005}, whereas FLDMA can achieve mutual independence even in single-path channels. Thus, FLDMA can provide full MIMO channel DoFs without relying on multipath effects for millimeter-wave and future higher-frequency communication systems.

\end{itemize}

The remainder of the paper is organized as follows. Section II introduces the FLDMA system model and provides the equivalent vectorization form. In Section III, the design of random permutation frequency offsets is proposed and the asymptotic orthogonality is proved. Meanwhile, the slight frequency offsets and pre-equalization methods are also discussed. The performance analysis of FLDMA is demonstrated in Section IV. Simulation results are provided in Section VI, and conclusions are drawn in Section VII.

\section{System Model}
We consider a wideband single-cell MU-MIMO communication system consisting of one BS equipped with $M$ antennas. The BS simultaneously provides services to $K$ UEs, each with $N$ antennas. The transmitter modules of both the BS and the UEs employ the FDA architecture. The channel between the BS and each UE is composed of $P+1$ propagation paths, comprising one Line-of-Sight (LoS) path and $P$ Non-Line-of-Sight (NLoS) paths. For simplicity in analysis, we assume that the MU-MIMO channel matrix adheres to the block fading condition, which implies that the complex-valued gains associated with each path remain constant throughout a data transmission block. Meanwhile, the Doppler shifts can be perfectly compensated at the receiving end. Furthermore, strict system synchronization is maintained across all devices within the cellular network, and the locations for BS and UEs are known to each other.


\subsection{FDA-MIMO-OFDM transceiver model}{\label{Section2A}}
We begin by introducing the single-user MIMO-OFDM channel model utilizing the FDA technique in the downlink system. Let $\mathbf{x}_{m}=\left[\mathrm{x}_{1,m},\mathrm{x}_{2,m},\dots,\mathrm{x}_{N_{\mathrm{us}},m}\right]^{T}$ be the information symbols sent by the BS on the $m$-th antenna. By employing OFDM modulation, the transmit signal of the BS can be expressed as
\begin{equation}
x_{m}(t) = \sum_{\ell=0}^{N_{\mathrm{us}}-1} \mathrm{x}_{\ell,m} g(t)e^{j2\pi \ell\Delta f t}, \quad -T_{\mathrm{cp}}\leq t < T_{\mathrm{us}},
\label{Eq2A.1}
 \end{equation}
where $\Delta f$ is the subcarrier spacing, $N_{\mathrm{us}}$ is the total number of subcarriers and $g(t)$ is the prototype pulse function. Then, the OFDM symbol duration is $T_{\mathrm{us}}=1/\Delta f$. Denote $N_{\mathrm{cp}}$ as the number of CPs, the CP length is equal to $T_{\mathrm{cp}}=N_{\mathrm{cp}}T_{\mathrm{us}}/N_{\mathrm{us}}$. Suppose that the $g(t)$ has unit energy and satisfies the following orthogonality
\begin{equation}
    \int g(t)g^{*}(t)e^{j2\pi (\ell -\ell^{\prime})\Delta f t} dt = \delta (\ell -\ell^{\prime}).
    \label{Eq2A.2}
\end{equation} 

Consider a generalized configuration for uniformly-spaced linear FDA made of $M$ antennas, where $\delta f_{m}$ is the frequency offset of mixing at the $m$-th antenna. Using ray-tracing based channel modeling approach, the overall signal arriving at $n$-th receive antenna of the $k$-th UE can be given by
\begin{equation}
\begin{aligned}
    y_{k,n}(t) = \frac{1}{\sqrt{MN}}\sum_{m=0}^{M-1}\sum_{p=0}^{P} &\sum_{\ell=0}^{N_{\mathrm{us}}-1} h_{k,p}\mathrm{x}_{\ell,m} g\left(t-\frac{R_{k,p,m,n}}{c}\right) \\
    & \cdot e^{j2\pi (f_{c} + \ell\Delta f+\delta f_{m}) \left(t-\tfrac{R_{k,p,m,n}}{c}\right)},
\label{Eq2A.3}
\end{aligned}
\end{equation}
where $f_{c}$ is center carrier frequency and the $h_{p}$ is the channel gain associated with the $p$-th propagation path. Concurrently, $R_{k,p,m,n}$ represents the propagation distance from the $m$-th transmit antenna to the $n$-th the receive antenna along the $p$-th path. $c$ is the speed of light. Here, the noise is omitted for brevity. Assuming both the scatterers and UEs are situated in the far-field regions, the $R_{k,p,m,n}$ can be calculated utilizing the planar-wave propagation model as follows
\begin{equation}
R_{k,p,m,n} = R_{k,p} - (m-1)d\sin{\theta_{k,p}}-(n-1)d\sin{\phi_{k,p}}.
\label{Eq2A.4}
\end{equation}
Taking the first element as the reference, $R_{k,p}$ represents the distance between the transmit reference element and the receive reference element for the $p$-th path. The element spacing between both the transmit and receive antennas is half the wavelength of the center carrier frequency, specifically $d=c/2f_{c}$. $\theta_{k,p}$ and $\phi_{k,p}$ respectively represent the angle-of-departure (AoD) and angle-of-arrival (AoA) in the $p$-th path. By substituting \eqref{Eq2A.4} into \eqref{Eq2A.3} and performing some manipulations, the received signal can be further expressed as
\begin{equation}
\begin{aligned}
        y_{k,n}(t) = &\frac{1}{\sqrt{MN}}\sum_{m=0}^{M-1}\sum_{p=0}^{P} \sum_{\ell=0}^{N_{\mathrm{us}}-1} \tilde{h}_{k,p} \mathrm{x}_{\ell,m} g\left(t-\frac{R_{k,p,m,n}}{c}\right) \\
        &\cdot e^{j2\pi(f_{c} + \ell\Delta f+\delta f_{m})\tfrac{(m-1)d\sin{\theta_{k,p}}+(n-1)d\sin{\phi_{k,p}}}{c}} \\ 
        &\cdot e^{j2\pi(\ell\Delta f+\delta f_{m})\left(t-\tfrac{R_{k}+r_{k,p}}{c}\right)}e^{j2\pi f_{c}t}.
        \label{Eq2A.5}
\end{aligned}
\end{equation}
In \eqref{Eq2A.5}, $\tilde{h}_{k,p}=h_{k,p}e^{j2\pi f_{c}R_{k,p}/c}$ is the complex gain for the $p$-th path. At the receiver, the center carrier $e^{j2\pi f_{c}t}$ can be eliminated by performing the down-conversion operation In this paper, we set $f_{c} \gg N_{\mathrm{us}}\Delta f+\delta f_{M-1}$, a condition that can be satisfied in the majority of communication systems. By this means, we have \cite{7522082}
\begin{equation}
\begin{aligned}
        &e^{j2\pi(f_{c} + \ell\Delta f+\delta f_{m})\tfrac{(m-1)d\sin{\theta_{k,p}}+(n-1)d\sin{\phi_{k,p}}}{c}} \\
        &\approx e^{j2\pi\tfrac{(m-1)\sin{\theta_{k,p}}+(n-1)\sin{\phi_{k,p}}}{2}}.
    \label{Eq2A.6}
\end{aligned}
\end{equation}
Simultaneously, this configuration allows us to suppose that all signals from the transmit antennas arrive at each receive antenna at nearly the same time, i.e., $R_{k,p,m,n} \approx R_{k,p}$. With the above approximations, \eqref{Eq2A.5} can be further simplified as
\begin{equation}
\begin{aligned}
    \bar{y}_{k,n}(t) \approx &\sum_{m=0}^{M-1}\sum_{p=0}^{P} \sum_{\ell=0}^{N_{\mathrm{us}}-1} \tilde{h}_{k,p}e^{-j2\pi \ell\Delta f \tfrac{R_{k,p}}{c}}\mathrm{x}_{\ell,m} \\
    &\cdot \mathbf{a}_{m}(R_{k,p},\theta_{k,p},t)\mathbf{b}_{n}(\phi_{k,p})
    g\left(t-\frac{R_{k,p}}{c}\right)e^{j2\pi\ell\Delta ft},
    \label{Eq2A.7}
\end{aligned}
\end{equation}
where $\mathbf{a}(R_{k,p},\theta_{k,p},t)$ represents the normalized FDA steering vector at the transmitter
\begin{equation}
\begin{aligned}
    \mathbf{a}(R_{k,p},\theta_{k,p},t) &= \frac{1}{\sqrt{M}}\left[1,e^{j2\pi\left(\tfrac{\sin{\theta_{k,p}}}{2}+\delta f_{1} \left(t-\tfrac{R_{k,p}}{c}\right)\right)}, \right. \\
    &\left. \cdots, e^{j2\pi\left(\tfrac{(M-1)\sin{\theta_{k,p}}}{2}+\delta f_{M-1}\left(t-\tfrac{R_{k,p}}{c}\right)\right)}\right]^{T}.
\label{Eq2A.8}
\end{aligned}
\end{equation}
From the above result, it becomes evident that the introduced series of frequency offsets in the spatial domain equips the array with distance-sensing capabilities. Therefore, UEs at different locations can be distinguished by combining angle and distance coordinates. The unique aspect of FDA lies in its time-varying steering vector, which necessitates the matched filters at the receiver to eliminate time dependency. 

Similarly, $\mathbf{b}(\phi_{k,p})$ represents the normalized PA steering vector at the receiver
\begin{equation}
    \mathbf{b}(\phi_{k,p}) = \frac{1}{\sqrt{N}}\left[1,e^{j2\pi\left(\tfrac{\sin{\phi_{k,p}}}{2}\right)},\cdots,e^{j2\pi\left(\tfrac{(N-1)\sin{\phi_{k,p}}}{2}\right)}\right]^{T}.
    \label{Eq2A.9}
\end{equation}

\begin{figure}[t]
    \centering
    \subfigure[Canonical method]{
        \begin{minipage}[t]{0.49\linewidth}
            \centering
            \includegraphics[width=1\linewidth]{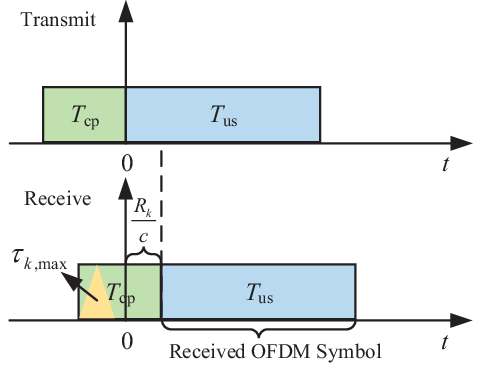}
        \end{minipage}%
    }%
    \subfigure[Proposed method]{
        \begin{minipage}[t]{0.49\linewidth}
            \centering
            \includegraphics[width=1\linewidth]{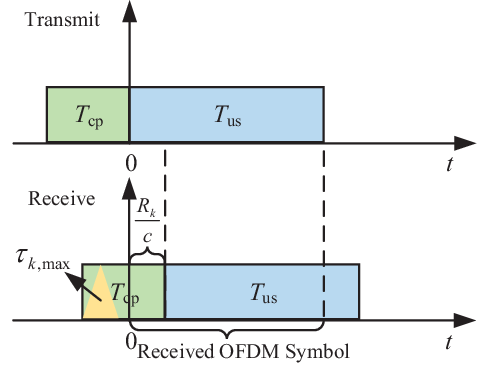}
        \end{minipage}%
    }%

    \centering
    \caption{Illustration of the CP removal method in our designed system.}
    \label{Fig.1}
\end{figure}

The canonical method discards the entire CP and the remaining signal is clipped for OFDM demodulation. In contrast, our designed receiver preserves signals within the interval $[0, T_{\mathrm{us}})$ under strict clock synchronization to retain distance information in the received signal. Consequently, a portion of the CP will be utilized in the subsequent signal processing, as shown in Fig. \ref{Fig.1}. Compared to neglecting the impact of propagation delay on pulse orthogonality, i.e., $\int g(t)g^{*}(t-\tau)e^{j2\pi \ell \Delta f (t-\tau))} dt \approx 0 \: \text{or} \: 1$, our approach ensures the integrity of all symbols in wideband systems. Therefore, it requires that the CP length should be greater than the maximum propagation delay of the channel, i.e. $T_{\mathrm{cp}} \geq R_{k,p}$, rather than the maximum multipath delay spread. Taking 15 kHz subcarrier as an example, 1/4 length CP could accommodate a maximum propagation path of 5 km, which is sufficient to meet the requirements of cellular communication. To retrieve the multiplexed symbols on each subcarrier, a key step is to employ $N_{\mathbf{us}}$ corresponding matched filters at the receiver. Considering $g(t)$ as a rectangular pulse, the received symbol on the $\ell^{\prime}$-th subcarrier can be expressed as \footnote{Distinguishing from existing FDA systems, where matched filters are typically designed as $g(t)\exp{\left\{j2\pi \delta f_{m} t\right\}}$ to mitigate the time-variance of beampattern. Here, we select $g(t)\exp{\left\{j2\pi \ell^{\prime}\Delta f t\right\}}$ as the matched filter signal for OFDM demodulation.}
\begin{equation}
\begin{aligned}
        &\mathrm{y}_{k,\ell^{\prime},n} = \int_{0}^{T_{\mathrm{us}}} \bar{y}_{k,n}(t)g^{*}(t)e^{-j2\pi \ell^{\prime} \Delta f t} dt \\
        &\overset{(a)}{\approx} \sum_{m=0}^{M-1} \sum_{\ell=0}^{N_{\mathrm{us}}-1} \underbrace{\sum_{p=0}^{P}\tilde{h}_{k,p}e^{-j2\pi \ell\Delta f \tfrac{R_{k,p}}{c}} \mathbf{a}_{m}(R_{k,p},\theta_{k,p})\mathbf{b}_{n}(\phi_{k,p})}_{\textstyle \mathbf{h}_{k,m,n,\ell}}    \\
        &\quad\quad \cdot\mathrm{x}_{\ell,m}\frac{1}{N_{\mathrm{us}}}\sum_{i=0}^{N_{\mathrm{us}}-1} e^{-j2\pi (\ell-\ell^{\prime}+\rho_{m})\tfrac{i}{N_{\mathrm{us}}}}\\
        & =\sum_{m=0}^{M-1} \alpha_{m}\mathbf{h}_{k,m,\ell^{\prime}}\mathrm{x}_{\ell^{\prime},m}+\sum_{m=0}^{M-1}\sum_{\ell=0,\ell\neq\ell^{\prime}}^{N_{\mathrm{us}}-1}\beta_{\ell,\ell^{\prime},m}\mathbf{h}_{k,m,n,\ell}\mathrm{x}_{\ell,m},
        \label{Eq2A.10}
\end{aligned}
\end{equation}
where approximation (a) is obtained by replacing summation with integral after sampling at intervals of $T_{\mathrm{us}}/N_{\mathrm{us}}$, $\mathbf{h}_{k,m,n,\ell}$ can be interpreted as the selective fading generated by multipath channels in the joint spatial and frequency domain and $\rho_{m}$ represents the ratio of frequency offset $\delta f_{m}$ to subcarrier spacing. The coefficients $\alpha_{m}$ and $\beta_{\ell,\ell^{\prime},m}$ are given by
\begin{subequations}
\begin{align}
\alpha_{m}&= {\rm Sa}_{N_{\mathrm{us}}}\left(\dfrac{\rho_{m}}{N_{\mathrm{us}}}\right) e^{j\pi\rho_{m}\tfrac{N_{\mathrm{us}}-1}{N_{\mathrm{us}}}},\\
\beta_{\ell,\ell^{\prime},m}&={\rm Sa}_{N_{\mathrm{us}}}\left(\dfrac{\ell-\ell^{\prime}+\rho_{m}}{N_{\mathrm{us}}}\right) e^{j\pi(\ell-\ell^{\prime}+\rho_{m})\tfrac{N_{\mathrm{us}}-1}{N_{\mathrm{us}}}},
\label{Eq2A.11}
\end{align}
\end{subequations}
where the function ${\rm Sa}_{N}(x)$ is defined as
\begin{equation}
    {\rm Sa}_{N}(x) = \dfrac{\sin{\pi Nx}}{N\sin{\pi x}}.
\end{equation}
From \eqref{Eq2A.10}, it can be observed that the frequency offset at each antenna disrupts the orthogonality of OFDM subcarriers, resulting in ICIs at the receiver. To address this, it is essential to design an equalization scheme that ensures perfect superposition of symbols on the same subcarrier across different antennas. On the other hand, frequency offsets spread the transmitting bandwidth, necessitating the use of a wideband receiver to capture the transmitted signal, which inevitably increases the noise power in the received symbols.

Let $\delta f_{m}=0, \forall m=0,1, \cdots, M-1 $, all ICI terms will vanish, and the model degenerates to 
\begin{equation}
    \mathrm{y}_{k,n,\ell}=\frac{1}{N_{\mathrm{us}}}\sum_{m=0}^{M-1}\sum_{p=0}^{P}\tilde{h}_{k,p}e^{-j2\pi \ell\Delta f \tfrac{R_{k,p}}{c}} \mathbf{a}_{m}(\theta_{k,p})\mathbf{b}_{n}(\phi_{k,p})\mathrm{x}_{\ell,m},
\end{equation}
which depicts a traditional MIMO-OFDM system relying on PA transceivers.

\subsection{Equivalent Vectorization Form}
\newcounter{mytempeqncnt}
\begin{figure*}[!t]
	\normalsize
	\setcounter{mytempeqncnt}{\value{equation}}
	\setcounter{equation}{15}
	\begin{equation}
    \begin{aligned}
   \bar{\mathbf{H}}_{k} &= \sum_{p=0}^{P} \tilde{h}_{k,p}\left(\mathbf{I}_{N} \otimes \mathbf{F}_{N_{\mathrm{us}}}\right) \left[\mathbf{b}(\phi_{k,p})\mathbf{a}^{T}(R_{k,p},\theta_{k,p}) \otimes \mathbf{I}_{N_{\mathrm{us}}}\right] \mathbf{\Xi}\left(\mathbf{I}_{M} \otimes \mathbf{F}^{H}_{N_{\mathrm{us}}} \right)\left(\mathbf{I}_{M}\otimes\mathbf{\Phi}_{k,p}\right) \\
   &= \sum_{p=0}^{P} \tilde{h}_{k,p} \underbrace{\left[\mathbf{b}(\phi_{k,p})\mathbf{a}^{T}(R_{k,p},\theta_{k,p}) \otimes \mathbf{F}_{N_{\mathrm{us}}} \right]\mathbf{\Xi}\left( \mathbf{I}_{M} \otimes \mathbf{F}^{H}_{N_{\mathrm{us}}} \mathbf{\Phi}_{k,p} \right)}_{\textstyle\bar{\mathbf{H}}_{k,p}},
    \label{Eq2B.3a}
    \end{aligned}
    \end{equation}
	\vspace*{4pt}

    \begin{equation}
        \bar{\mathbf{H}}_{k,p}=\left[\begin{matrix}
            &b_{0}a_{0}\mathbf{F}_{N_{\mathrm{us}}}\mathbf{\Xi}_{0} \mathbf{F}^{H}_{N_{\mathrm{us}}}\mathbf{\Phi}_{k,p} &b_{0}a_{1}\mathbf{F}_{N_{\mathrm{us}}}\mathbf{\Xi}_{1} \mathbf{F}^{H}_{N_{\mathrm{us}}}\mathbf{\Phi}_{k,p} & \cdots & b_{0}a_{M-1}\mathbf{F}_{N_{\mathrm{us}}}\mathbf{\Xi}_{M-1} \mathbf{F}^{H}_{N_{\mathrm{us}}}\mathbf{\Phi}_{k,p} \\
    &b_{1}a_{0}\mathbf{F}_{N_{\mathrm{us}}}\mathbf{\Xi}_{0} \mathbf{F}^{H}_{N_{\mathrm{us}}}\mathbf{\Phi}_{k,p} &b_{1}a_{1}\mathbf{F}_{N_{\mathrm{us}}}\mathbf{\Xi}_{1} \mathbf{F}^{H}_{N_{\mathrm{us}}}\mathbf{\Phi}_{k,p} & \cdots & b_{1}a_{M-1}\mathbf{F}_{N_{\mathrm{us}}}\mathbf{\Xi}_{M-1} \mathbf{F}^{H}_{N_{\mathrm{us}}}\mathbf{\Phi}_{k,p} \\
    &\vdots & \vdots & \ddots & \vdots \\
    &b_{N-1}a_{0}\mathbf{F}_{N_{\mathrm{us}}}\mathbf{\Xi}_{0} \mathbf{F}^{H}_{N_{\mathrm{us}}}\mathbf{\Phi}_{k,p} &b_{N-1}a_{1}\mathbf{F}_{N_{\mathrm{us}}}\mathbf{\Xi}_{1} \mathbf{F}^{H}_{N_{\mathrm{us}}}\mathbf{\Phi}_{k,p} & \cdots & b_{N-1}a_{M-1}\mathbf{F}_{N_{\mathrm{us}}}\mathbf{\Xi}_{M-1} \mathbf{F}^{H}_{N_{\mathrm{us}}}\mathbf{\Phi}_{k,p} 
        \end{matrix}\right],
        \label{Eq2B.3b}
    \end{equation}
	\hrulefill
	\setcounter{equation}{16}
	\vspace*{4pt}
\end{figure*}
Since the MIMO system can be more compactly represented in vectorization form, the transmitted and received symbols are taken as column vectors for the convenience of the following analysis. Because of the intrinsic ICIs yielded in the FDA-MIMO-OFDM system, it is necessary to consider the input-output relationship of all information symbols within an OFDM symbol duration. 

Let us define $\mathbf{x} \in \mathbb{C}^{N_{\mathrm{us}}M\times 1}$ as the vectorization form of $x_{\ell,m}$, where its $(mN_{\mathrm{us}} + \ell)$-th entry is $x_{\ell,m}$, and $\mathbf{y}_{k} \in \mathbb{C}^{N_{\mathrm{us}}N\times 1}$ as the vectorization form of $y_{k,\ell^{\prime},n}$, where its $(nN_{\mathrm{us}} + \ell^{\prime})$-th entry is $y_{k,\ell^{\prime},m}$. Based on the system introduced above, an equivalent form of the FDA-MIMO-OFDM transceiver is given as follows
\begin{equation}
    \mathbf{y}_{k} = \sum_{p=0}^{P} \left(\mathbf{I}_{N} \otimes \mathbf{F}_{N_{\mathrm{us}}}\right) \mathbf{H}_{k,p} \left(\mathbf{I}_{M} \otimes \mathbf{F}^{H}_{N_{\mathrm{us}}} \right)\left(\mathbf{I}_{M}\otimes\mathbf{\Phi}_{k,p}\right)\mathbf{x},
    \label{Eq2B.1}
    	\setcounter{equation}{14}
\end{equation}
where $\mathbf{F}_{N_{\mathrm{us}}}$ denotes the $N_{\mathrm{us}}$-point discrete-Fourier-transform matrix, $\otimes$ denotes Kronecker product, $\mathbf{\Phi}_{k,p}=\mathrm{diag} \left(1,e^{-j2\pi\Delta f R_{k,p}/c},\cdots,e^{-j2\pi(N_{\mathrm{us}}-1)\Delta f R_{k,p}/c}\right)$ is the phase shifts caused by the propagation delay on each subcarrier, and $\mathbf{H}_{k,p}\in \mathbb{C}^{N_{\mathrm{us}}N \times N_{\mathrm{us}}M}$ is the space-time domain channel matrix including the effect of $M$ frequency offsets for the $p$-th path, which can be written as
\begin{equation}
    \mathbf{H}_{k,p} = \tilde{h}_{k,p}\left[\mathbf{b}(\phi_{k,p})\mathbf{a}^{T}(R_{k,p},\theta_{k,p}) \otimes \mathbf{I}_{N_{\mathrm{us}}}\right]\mathbf{\Xi}.
\label{Eq2B.2}
\end{equation}
where the entries in $\mathbf{\Xi} = \mathrm{diag}\left(\mathbf{\Xi}_{0},\mathbf{\Xi}_{1},\cdots,\mathbf{\Xi}_{M-1}\right)$ represent the phase shifts caused by frequency offsets at each sampling instant, where $\mathbf{\Xi}_{m} = \mathrm{diag}\left(1,e^{j2\pi \rho_{m}/N_{\mathrm{us}}}, \cdots,e^{j2\pi \rho_{m}(N_{\mathrm{us}}-1)/N_{\mathrm{us}}}\right)$. By using the Kronecker product rule, the space-frequency domain channel matrix for the vectorized channel input $\mathbf{x}$ and output $\mathbf{y}_{k}$ for the $k$-th UE can be obtained by \eqref{Eq2B.3a}, where $\bar{\mathbf{H}}_{k,p}$ is the space-frequency domain channel matrix of the $p$-path, which can be further explicitly given by \eqref{Eq2B.3b}, where $a_{m}$ and $b_{m}$ represent the $m$-th and $n$-th elements in $\mathbf{a}(R_{k,p},\theta_{k,p})$ and $\mathbf{b}(\phi_{k,p})$, respectively. 

Analogously, let $\delta f_{m}=0, \forall m=0,1, \cdots, M-1 $, the matrix $\mathbf{\Xi}_{m}$ will degenerate to $\mathbf{I}_{N_{\mathrm{us}}}$, and the $\bar{\mathbf{H}}_{k,p}$ will evolve into a block diagonal matrix. Consequently, the FDA transceiver model is reduced to the following PA transceiver model
\begin{equation}
    \mathbf{y}_{k,\ell} =  \sum_{p=0}^{P} \tilde{h}_{k,p}\mathbf{b}(\phi_{k,p}) \mathbf{a}^{T}(\theta_{k,p})\mathbf{\Phi}_{k,p}(\ell) \mathbf{x}_{\ell},
    \label{Eq2B.4}
    \setcounter{equation}{18}
\end{equation}
where $\mathbf{y}_{k,\ell}\in\mathbb{C}^{N\times1}, \mathbf{x}_{\ell}\in\mathbb{C}^{M\times 1}$ represent the received and transmitted vectors on the $\ell$-th OFDM subcarrier. The effective vectorized input-output relationship will be utilized in the MU-MIMO model and subsequent system performance analysis.


\subsection{MU-MIMO Model}
Consider a downlink MU-MIMO system, $\mathbf{s} =\left[(\mathbf{s} _{0})^{T},(\mathbf{s} _{1})^{T},\cdots,(\mathbf{s}_{K-1})^{T}\right]^{T}$ is the overall information vector, where $\mathbf{s}_{k}=\left[s_{k,0},s_{k,1},\cdots,s_{k, N_{\mathrm{us}}-1}\right]^{T}$ is the quadrature amplitude modulation (QAM) symbol vector sent to $k$-th UE, satisfying $\mathbf{E} \left[| \mathrm{s}_{k,\ell}|^{2} \right]=1$. Then, the transmit signal is given by $\mathbf{x}  =  \mathbf{W} \mathbf{A} \mathbf{s} $, where $\mathbf{A}  = \mathrm{diag} \left(a_{0}  \mathbf{I}_{N_\mathrm{us}}, a_{1} \mathbf{I}_{N_\mathrm{us}}, \cdots, a_{K-1}  \mathbf{I}_{N_\mathrm{us}}\right)$ is the power normalisation matrix, $a_{k}$ is the transmit power allocated to $k$-th UE. The transmit precoding matrix $ \mathbf{W}$ has been extended to $N_{\mathrm{us}}M\times N_{\mathrm{us}}K$ dimension in the FDA system, given by $ \mathbf{W} = \left[ \mathbf{W}_{0},  \mathbf{W}_{1}, \cdots, \mathbf{W}_{K-1}\right]$. Concretely, $ \mathbf{W}_{k}= \mathbf{w}_{k}\otimes \mathbf{I}_{N_\mathrm{us}}$, where $ \mathbf{w}_{k}\in \mathbb{C}^{M\times 1}$ is the normalized transmit precoding vector of $k$-th UE satisfing $\left\| \mathbf{w}_{k}\right\|^{2}=1$. The composite channel matrix is given by stacking the space-frequency domain channel matrics on top of each other, given by $\bar{\mathbf{H}}  = \left[\bar{\mathbf{H}} _{0}, \bar{\mathbf{H}} _{1}, \cdots,\bar{\mathbf{H}} _{K-1}\right]^{T}\in\mathbb{C}^{N_{\mathrm{us}}NK\times N_{\mathrm{us}}M}$. The additive noise $\mathbf{n}_{k}\in\mathbb{C}^{N_{\mathrm{us}}N\times 1}$ follows an independent complex Gaussian variable with zero mean and equal variance $\sigma^{2}_{n}$ for all users, i.e., $\mathbf{n}_{k} \sim \mathcal{CN} \left(\mathbf{0}, \sigma^{2}_{n} \mathbf{I}_{N_{\mathrm{us}}N}\right)$. The received signal at $k$-th can be written as
\begin{equation}
    \mathbf{y} _{k} = \bar{\mathbf{H}} _{k} \mathbf{W}_{k} a _{k} \mathbf{s} _{k} + \bar{\mathbf{H}} _{k}\sum_{i \neq k}  \mathbf{W}_{i} a _{i} \mathbf{s} _{i} + \mathbf{n} _{k},
        \label{Eq2C.1}
\end{equation}
and the received signal for all $K$ UEs can be represented as
\begin{equation}
    \mathbf{y}  = \bar{\mathbf{H}}  \mathbf{W} \mathbf{A}  \mathbf{s}  + \mathbf{n} ,
            \label{Eq2C.2}
\end{equation}
where $\mathbf{n} \in\mathbb{C}^{N_{\mathrm{us}}NK\times 1}$ is the stacked Gaussian noise vector.

\section{Waveform Design}
From the results in the previous section, it can be observed that frequency offsets endow the perception of distance dimension to the steering vector while disrupting the orthogonality of the OFDM transmission system. Additionally, frequency offsets also determine the orthogonality of the beams in the distance-angle plane. To enhance orthogonal transmission in communication systems, we investigate waveform design schemes based on random permutation frequency offsets in this section. By analyzing the correlation between different steering vectors, the asymptotic orthogonality in distance and angle domain is revealed. Subsequently, we discuss design approaches for orthogonal transmission FLDMA systems from two aspects: employing slight frequency offsets or pre-equalization.

\subsection{Random Permutation Frequency Offsets}
\begin{figure}[tbp]
    \centering
    \subfigure[PA]{
        \begin{minipage}[t]{0.49\linewidth}
            \centering
            \includegraphics[width=1\linewidth]{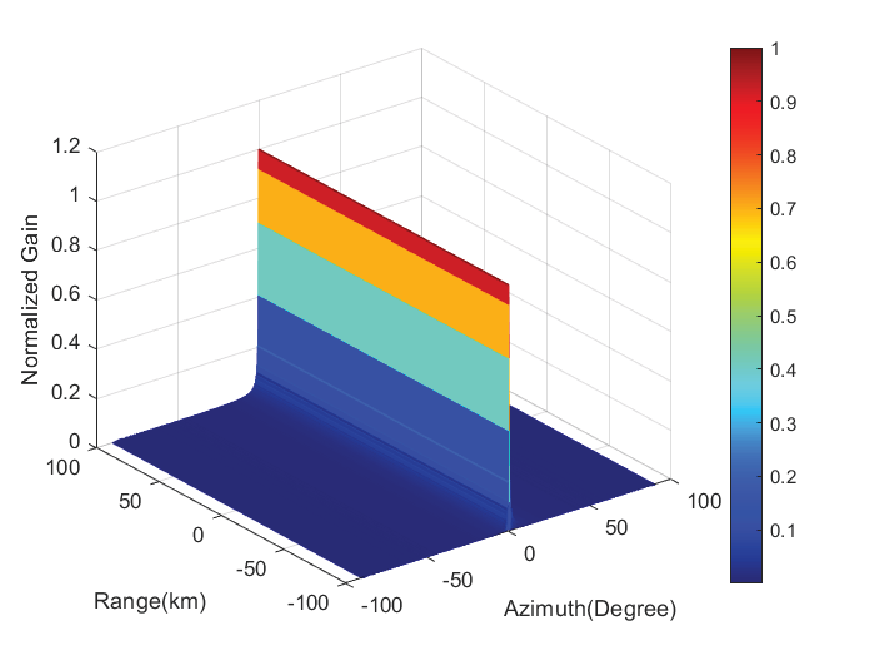}
            \label{Fig.3a}
        \end{minipage}%
    }%
    \subfigure[FDA with uniformly increased frequency offsets]{
        \begin{minipage}[t]{0.49\linewidth}
            \centering
            \includegraphics[width=1\linewidth]{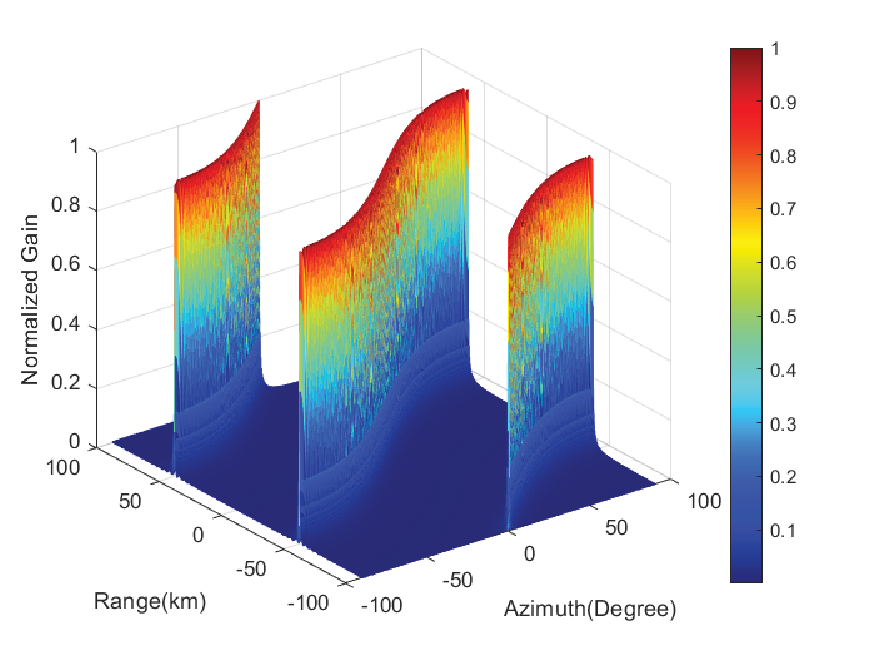}
            \label{Fig.3b}
        \end{minipage}%
    }  %
    
    \subfigure[FDA with logarithmically increased frequency offsets]{
        \begin{minipage}[t]{0.49\linewidth}
            \centering
            \includegraphics[width=1\linewidth]{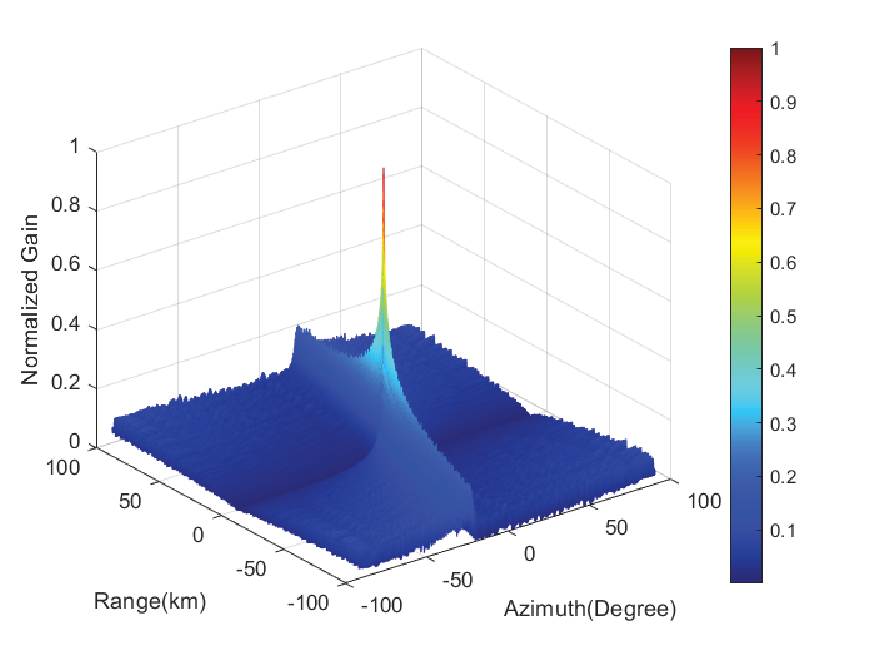}
            \label{Fig.3c}
        \end{minipage}%
    }%
    \subfigure[FDA with random permutation frequency offsets]{
        \begin{minipage}[t]{0.49\linewidth}
            \centering
            \includegraphics[width=1\linewidth]{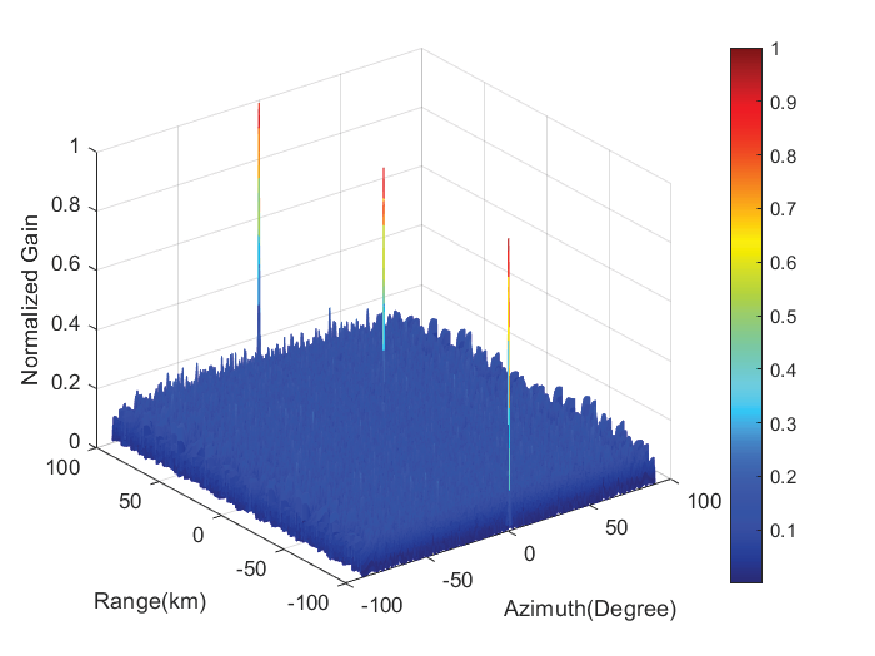}
            \label{Fig.3d}
        \end{minipage}%
    }  %
    \centering
    \caption{Comparison of beampattern between PA and different FDAs, where $M=256$, $\delta f=3$ kHz. To ensure the same maximum frequency offset, the logarithmically increased frequency offsets are set to $\delta f_{m}=\left(M-1\right)\ln\left(m+1\right)\delta f/\ln M $.}
    \label{Fig.3}
\end{figure}

In the original FDA, frequency offsets are uniformly increased, i.e., $\delta f_{m} = (m-1)\delta f$, where $\delta f$ is the basic frequency increment cell. Based on the FDA steering vector in an equivalent baseband model, the correlation of two beam vectors corresponding to the locations of $\left(R_{i},\theta_{i}\right)$ and $\left(R_{j},\theta_{j}\right)$ can be formulated as
\begin{equation}
\begin{aligned}
    &\left|\mathbf{a}^{H}\left(R_{i},\theta_{i}\right)\mathbf{a}\left(R_{j},\theta_{j}\right)\right|\\
     &  \quad \quad= \frac{1}{M}\left|\sum_{m=0}^{M-1}e^{j2\pi m\left(\frac{\sin{\theta_{i}}-\sin{\theta_{j}}}{2}-\delta f\frac{R_{i}-R_{j}}{c}\right)}\right| \\
     &\quad \quad= {\rm Sa}_{M}\left(\frac{\sin{\theta_{i}}-\sin{\theta_{j}}}{2}-\delta f\frac{R_{i}-R_{j}}{c}\right).
     \label{Eq3A.1}
\end{aligned}
\end{equation}
From \eqref{Eq3A.1}, the correlation of the steering vectors will achieve the maximum when $\left(\sin{\theta_{i}}-\sin{\theta_{j}}\right)/2-\delta f\left(R_{i}-R_{j}\right)/c=l, l\in \mathbb{Z}$. It can be observed that the distance and angle within the peak region exhibit a sinusoidal relationship, resulting in an "S"-shaped beampattern as shown in Fig. \ref{Fig.3b}. The primary challenge posed by this "S"-shaped beam lies in the mutual coupling between distance and angle, which can lead to multiple pairs of distance and angle solutions for a target. This ambiguity prevents the beam from precisely focusing on a specific UE. if two UEs fall within the "S"-shaped beam region, they will experience the identical channel, making it impossible for the BS to differentiate between them in the spatial domain.

To generate dot-shaped beams, we first elucidate the underlying cause of the coupling between distance and angle dimensions. From \eqref{Eq3A.1}, we observe that the correlation peaks only when the phases of all terms in the summation are aligned. This requires that $\sin{\theta_{i}}-\sin{\theta_{j}}$ and $R_{i}-R_{j}$ need to satisfy the following equation
\begin{equation}
    \left[ \mathbf{G}_{1}, \mathbf{G}_{2}\right]\left[\sin{\theta_{i}}-\sin{\theta_{j}},R_{i}-R_{j}\right]^{T}=\left[0,l_{1},\cdots,l_{M-1}\right]^{T},
     \label{Eq3A.2}
\end{equation}
where $ \mathbf{G}_{1}=\left[0,1/2,\cdots,\left(M-1\right)/2\right]^{T}$, $ \mathbf{G}_{2}=\left[0,-\delta f/c,\cdots,-\left(M-1\right)\delta f/c\right]^{T}$ and $l_{1},\cdots,l_{M-1} \in \mathbb{Z}$. Since $ \mathbf{G}_{1}$ and $ \mathbf{G}_{2}$ are linearly dependent, there exist infinitely many solutions for $\left(R_{j},\theta_{j}\right)$ corresponding to a fixed $\left(R_{i},\theta_{i}\right)$, leading to the observed coupling between the distance and angle dimensions \cite{https://doi.org/10.1049/el.2017.2355}. Therefore, breaking the dependency between $ \mathbf{G}_{1}$ and $ \mathbf{G}_{2}$ is essential for generating focused beams in FDA. A straightforward and effective method is to randomly shuffle the elements in $ \mathbf{G}_{2}$ by shuffling $\delta f_{m}$, which can be regarded as a special implementation of random frequency offsets. Fig. \ref{Fig.3} illustrates the beampatterns of various frequency offset configurations. Among these, the random permutation frequency offsets produce nearly dot-shaped beams, demonstrating superior two-dimensional directivity compared to logarithmically increasing frequency offsets. This outcome indicates a successful decoupling of the distance and angle dimensions. Moreover, this frequency offset configuration preserves the fast Fourier transform (FFT) properties in the distance domain, mirroring those in the angle domain. Consequently, the beampatterns of both uniformly increased frequency offsets and random permutation frequency offsets exhibit periodicity. Based on the FFT properties, the beamwidth in the distance domain is $c/\left[\left(M-1\right)\delta f\right]$, with the corresponding period being $c/\delta f$. 

According to \eqref{Eq3A.1}, we can easily obtain the clarified expression of correlation between two steering vectors at the same distance or angle. Due to the randomness of frequency offsets, precise computation of the correlation is unfeasible when $R_{i}\neq R_{j}$ and $\theta_{i} \neq \theta_{j}$. However, employing statistical properties allows for the analysis of the correlations between beams to some extent. Considering the correlation as a stochastic process with respect to distance and angle, it can be characterized by the following lemma.

\begin{lemma}\label{lemma1}
Let $\delta f_{m}=z_{m}\delta f$, where $z_{m}=\boldsymbol{\pi}_{m}\left(\mathcal{M}\right)$ denotes the $m$-th element in randomly shuffled sequence $\mathcal{M}=[0,1,\cdots, M-1]$. Then, the mean and variance of beams correlation $\eta_{i,j} = \mathbf{a}^{H} \left(R_{i},\theta_{i}\right) \mathbf{a} \left(R_{j}, \theta_{j}\right)$ are given by
\begin{subequations}
    \begin{equation}
            \mathbf{E} \left[\eta_{i,j} \right] = {\rm Sa}_{M}\left(p_{i,j}\right){\rm Sa}_{M}\left(q_{i,j}\right)e^{j\pi\frac{M-1}{M}\left(p_{i,j}-q_{i,j}\right)},
    \end{equation}
    \begin{equation}
    \begin{aligned}
        \mathbf{Var}\left[\eta_{i,j}\right] = &\dfrac{M}{M-1} \left({\rm Sa}_{M}^{2}\left(p_{i,j}\right)-\frac{1}{M}\right)\left({\rm Sa}_{M}^{2}\left(q_{i,j}\right) -\frac{1}{M}\right) \\
        &-{\rm Sa}_{M}^{2}\left(p_{i,j}\right){\rm Sa}^{2}_{M}\left(q_{i,j}\right)+\frac{1}{M},
    \end{aligned}
    \end{equation}
\end{subequations}
where we denote $p_{i,j}=(\sin{\theta_{i}}-\sin{\theta_{j}})/2$ and $q_{i,j}=\delta f \left(R_{i}-R_{j}\right)/c$.
\end{lemma}
\begin{myproof}
The beam correlation using random permutation frequency offsets can be calculated as
\begin{equation}
\begin{aligned}
        \eta_{i,j} &= \frac{1}{M}\sum_{m=0}^{M-1}e^{j2\pi \left[m\left(\frac{\sin{\theta_{i}}-\sin{\theta_{j}}}{2}\right)-z_{m}\left(\delta f\frac{R_{i}-R_{j}}{c}\right)\right]}.
        \label{EqA.1}
\end{aligned}
\end{equation}
Since $z_{m}$ is a random sampling variable, the mean of $\eta\left(q_{i,j},p_{i,j}\right)$ can be calculated as
\begin{equation}
    \begin{aligned}
        &\mathbf{E} \left[\eta\left(q_{i,j},p_{i,j}\right) \right] \\
        &= \int_{\mathbf{z}\in\mathcal{M}^{M}} p(\mathbf{z}) \eta\left(q_{i,j},p_{i,j}\right) d \mathbf{z} \\
        &= \int_{\mathbf{z}\in\mathcal{M}^{M}}  p(z_{m} \mid\mathbf{z} \backslash z_{m})p(\mathbf{z} \backslash z_{m})\eta\left(q_{i,j},p_{i,j}\right) d \mathbf{z} \\
        &= \frac{1}{M}\sum_{m=0}^{M-1}e^{j2\pi mq_{i,j}} \int_{\mathbf{z}\in\mathcal{M}^{M}}  p(z_{m} \mid\mathbf{z} \backslash z_{m})p(\mathbf{z} \backslash z_{m}) \\
        & \quad \quad \cdot e^{-j2\pi z_{m}p_{i,j}}dz_{m} d\mathbf{z} \backslash z_{m} \\
        & \overset{(a)}{=} \frac{1}{M}\sum_{m=0}^{M-1}e^{j2\pi mq_{i,j}} \frac{1}{M}\sum_{m=0}^{M-1}e^{j2\pi z_{m}p_{i,j}} \\
        &= {\rm Sa}_{M}(q_{i,j}){\rm Sa}_{M}(p_{i,j})e^{j\pi\frac{M-1}{M}\left(p_{i,j}-q_{i,j}\right)}.
        \label{EqA.2}
    \end{aligned}
\end{equation}
where $\mathbf{z} \backslash z_{m}$ represents all the elements of $\mathbf{z}$ outside $z_{m}$. For a randomly shuffled sequence, it holds that $p(z_{m} \mid\mathbf{z} \backslash z_{m})=1$ and $p(\mathbf{z} \backslash z_{m})=1/M!$. Therefore, equation (a) can be obtained. 
The variance of $\eta\left(q_{i,j},p_{i,j}\right)$ can be obtained by
\begin{equation}
\begin{aligned}
    &\mathbf{Var}\left[\eta_{i,j}\right] \\
    &= \mathbf{E}\left[ \left|\eta_{i,j}\right|^{2}\right]-\dfrac{1}{M^{2}}\left|\mathbf{E}\left[\eta_{i,j}\right]\right|^{2} \\
    &=\dfrac{1}{M} + \sum_{m=0}^{M-1}\sum_{n=0,n\neq m}^{M-1} e^{j\pi\left(m-n\right)p_{i,j}} \mathbf{E}\left[ e^{j\pi\left(z_{m}-z_{n} \right)q_{i,j}}\right] \\
    & \quad- \left|\mathbf{E}\left[\eta_{i,j}\right]\right|^{2},
    \label{EqA.3}
    \end{aligned}
\end{equation}
where 
\begin{equation}
    \begin{aligned}
        &\mathbf{E}[e^{ j\pi\left(z_{m}-z_{n} \right)q_{i,j}}]  \\
        &= \int_{\mathbf{z}_{m}\in\mathcal{M}^{M},\mathbf{z}_{n}\in\left(\mathcal{M}\backslash {z}_{m}\right)^{M-1}} p(\mathbf{z}_{m})p(\mathbf{z}_{n}) \\
        & \quad \cdot e^{j\pi\left(z_{m}-z_{n} \right)q_{i,j}} d\mathbf{z}_{m} d\mathbf{z}_{n} \\
       & = \int_{\mathbf{z}_{n}\in\mathcal{M}^{M}} \dfrac{1}{M!} d\mathbf{z}_{m} \backslash z_{m} \int_{\mathbf{z}_{n}\in\left(\mathcal{M}\backslash {z}_{m}\right)^{M-1}} \frac{1}{\left(M-1\right)!} d\mathbf{z}_{n} \backslash z_{n} \\
       &\quad \cdot \sum_{m=0}^{M-1}\sum_{n=0,n\neq m}^{M-1} e^{j\pi\left(z_{m}-z_{n} \right)q_{i,j}} \\
       & =\frac{1}{M}\frac{1}{\left(M-1\right)}\left(\sum_{m=0}^{M-1}\sum_{n=0}^{M-1} e^{j\pi\left(z_{m}-z_{n} \right)q_{i,j}} - M\right) \\
       &= \frac{M}{M-1} \left({\rm Sa}_{M}^{2}(q_{i,j})-\frac{1}{M}\right).
    \label{EqA.4}
    \end{aligned}
\end{equation}
Substituting \eqref{EqA.4} into \eqref{EqA.3}, the $\mathbf{Var}\left[\eta_{i,j}\right]$ can be further rewritten as
\begin{equation}
\begin{aligned}
        \mathbf{Var}\left[\eta_{i,j}\right] = &\dfrac{M}{M-1} \left({\rm Sa}_{M}^{2}\left(p_{i,j}\right)-\frac{1}{M}\right)\left({\rm Sa}_{M}^{2}\left(q_{i,j}\right) -\frac{1}{M}\right) \\
        &-{\rm Sa}_{M}^{2}\left(p_{i,j}\right){\rm Sa}^{2}_{M}\left(q_{i,j}\right)+\frac{1}{M}.
        \end{aligned}
\end{equation}
This completes the proof.
\end{myproof}

\begin{figure}[tbp]
    \centering
    \subfigure[Simulation result of mean value]{
        \begin{minipage}[t]{0.49\linewidth}
            \centering
            \includegraphics[width=1\linewidth]{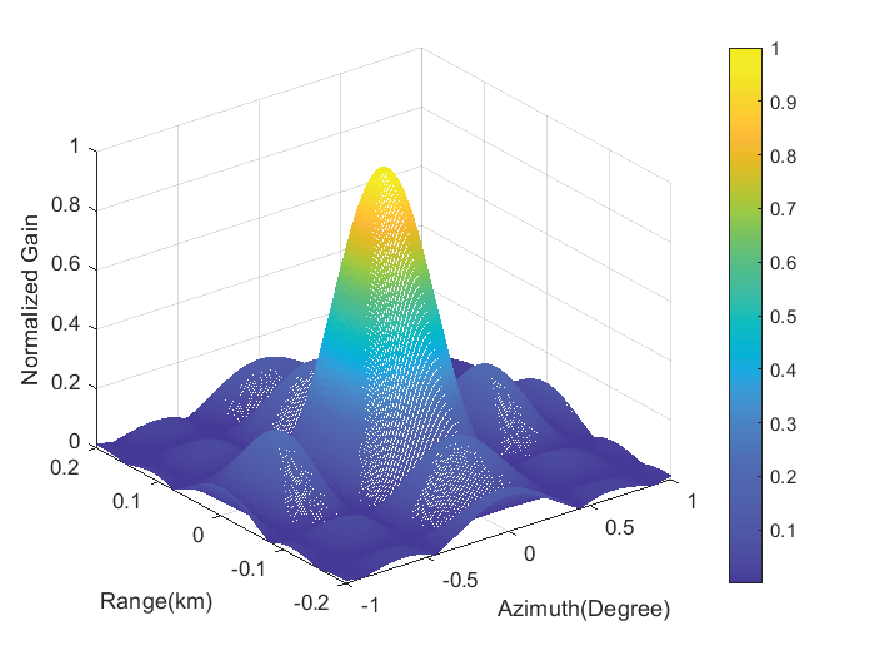}
            \label{Fig.4a}
        \end{minipage}%
    }%
    \subfigure[Theoretical result of mean value]{
        \begin{minipage}[t]{0.49\linewidth}
            \centering
            \includegraphics[width=1\linewidth]{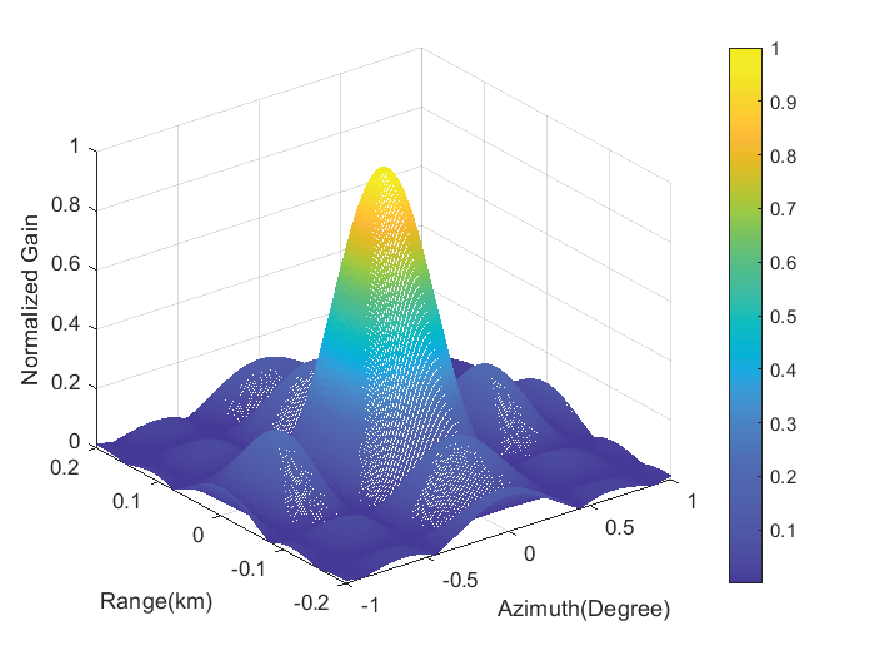}
            \label{Fig.4b}
        \end{minipage}%
    }  %
    
    \subfigure[Simulation result of variance value]{
        \begin{minipage}[t]{0.49\linewidth}
            \centering
            \includegraphics[width=1\linewidth]{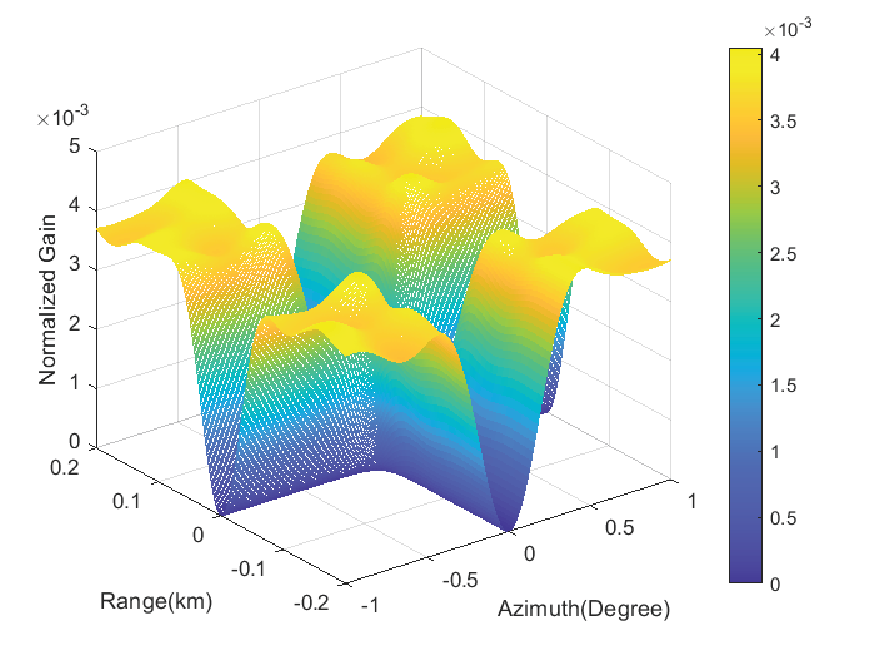}
            \label{Fig.4c}
        \end{minipage}%
    }%
    \subfigure[Theoretical result of variance value]{
        \begin{minipage}[t]{0.49\linewidth}
            \centering
            \includegraphics[width=1\linewidth]{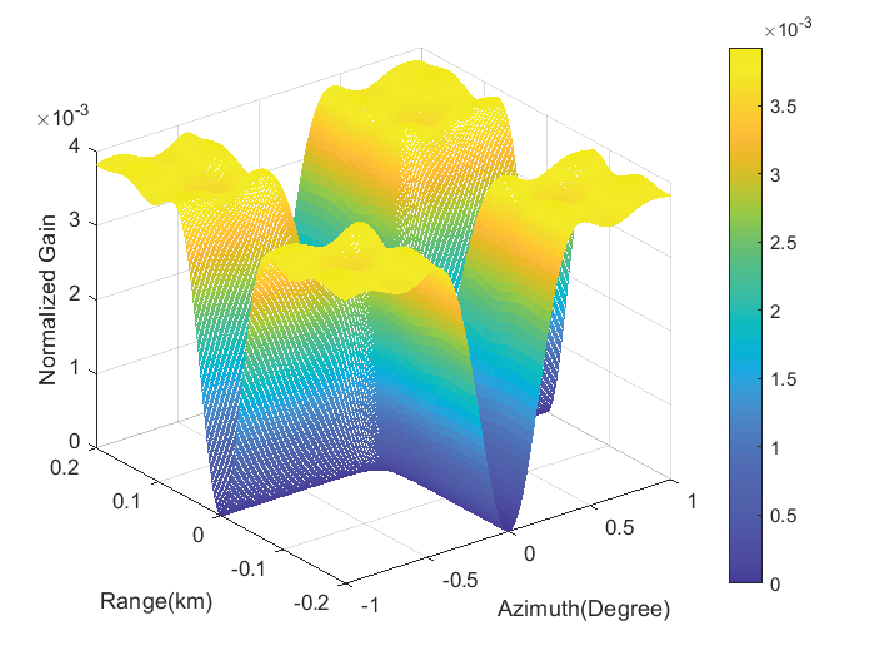}
            \label{Fig.4d}
        \end{minipage}%
    }  %
    \centering
    \caption{The mean and variance results of beam correlation.}
    \label{Fig.4}
\end{figure}

To verify the accuracy of the moments of correlation, we illustrate a comparison between the theoretical results and the simulation results in Fig. \ref{Fig.4}. All the simulations in the left column are counted from 1000 Monte Carlo trials, demonstrating a well match with the theoretical derivation. Then, the correlation can be approximated to 
\begin{equation}
\begin{aligned}
            &\left|\eta_{i,j}\right|^{2} \\
            &\approx \dfrac{1}{M} +\dfrac{M}{M-1} \left({\rm Sa}_{M}^{2}\left(p_{i,j}\right)-\frac{1}{M}\right)\left({\rm Sa}_{M}^{2}\left(q_{i,j}\right) -\frac{1}{M}\right).
        \label{Eq3A.3}
\end{aligned}
\end{equation}
Based on the above analysis, we can prove the asymptotic orthogonality in the distance-angle domain in the following corollary.

\begin{corollary}{\label{corollary1}}
As the number of antennas increases, the steering vectors corresponding to any angle or distance in the FDA with random permutation frequency offsets approach asymptotically become orthogonal within one distance period.
\begin{equation} 
\lim_{M\rightarrow+\infty}\left|\mathbf{a}^{H}\left(R_{i},\theta_{i}\right)\mathbf{a}\left(R_{j},\theta_{j}\right)\right| =0,  \text{for} \ R_{i}\neq R_{j} \ \text{or} \ \theta_{i} \neq \theta_{j}.
\label{Eq3A.4}
\end{equation}
\end{corollary}

\begin{myproof}
According to \eqref{Eq3A.3}, as the number of antennas tends to infinity, the mean and variance of beta approach 0 when $R_{i}\neq R_{j}$ or $\theta_{i} \neq \theta_{j}$. Therefore, the correlation $\lim_{M\rightarrow+\infty} \left|\eta_{i,j}\right| =0$, indicating asymptotic orthogonality of the steering vectors in the distance-angle domain. This completes the proof.
\end{myproof}

This corollary indicates that the FDA system with random permutation frequency offsets attains 2D orthogonality in the distance-angle domain, whereas the PA system achieves orthogonality solely in the angle domain. 

Although FLDMA demonstrates superior performance in the spatial domain compared to SDMA, this advantage comes at the cost of generating ICIs, which results in a severe imbalance in the singular values corresponding to each subcarrier.  Furthermore, the additional spectral overhead and increased noise power exacerbate the performance degradation. Consequently, the potential of FLDMA to enhance spectral efficiency in MIMO systems requires further investigation.

\begin{figure*}[ht!]
    \centering
    \includegraphics[width=6in]{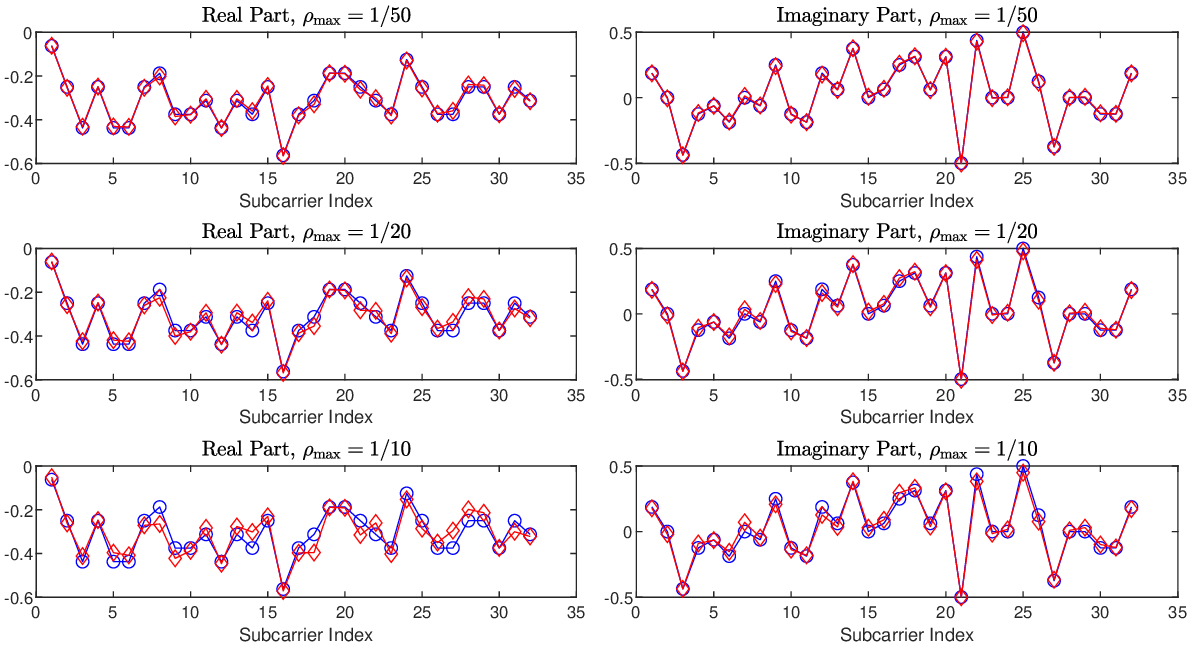}
    
    

    \centering
    \caption{The mean and variance results of beam correlation.}
    \label{Fig.5}
\end{figure*}
\subsection{Slight Frequency Offsets}

While the FDA can satisfy the 2D asymptotic orthogonality of steering vectors, the trade-off is the introduction of ICIs as described in \eqref{Eq2A.10}. We can observe that not only the phase of the received symbol is shifted, but also the interferences occur among distinct subcarriers across various antennas. A straightforward approach is to mitigate interference by employing slight frequency offsets to ensure the following condition
\begin{equation}
    \int_{0}^{T_{\mathrm{us}}} \left|g(t)\right|^{2}\exp{\left\{-j2\pi (\ell^{\prime}-\ell + \rho_{m}) \Delta f t\right\}} dt \approx  
    \begin{cases}
        1, \quad \ell^{\prime} \neq \ell, \\
        0,  \quad \ell^{\prime} = \ell,
    \end{cases}    
    \label{Eq3B.1}
\end{equation}
so that the received symbol on the $\ell^{\prime}$-th subcarrier can be reduced to $y_{k,\ell^{\prime},n} =\sum_{m=0}^{M-1} \mathbf{h}_{k,m,\ell^{\prime}}\mathrm{x}_{\ell^{\prime},m}$. The equivalent matrix form of \eqref{Eq3B.1} can be written as $\mathbf{F}_{N_{\mathrm{us}}} \mathbf{\Xi} \mathbf{F}^{H}_{N_{\mathrm{us}}} \approx \mathbf{I}_{N_{\mathrm{us}}}$, thus the system is capable of approximate orthogonal transmission as \eqref{Eq2B.4}. Considering the assumption that UEs are aware of the precise location of the BS, UE can compensate for the phase shift $\exp{\left(-j2\pi \ell \Delta f R_{k,0}/c\right)}$ on each subcarrier in the received signal. As a result, it yields the ideal channel matrix $\bar{\mathbf{H}}^{\text{ideal}} = \mathbf{b}(\phi) \mathbf{a}^{T}(R,\theta) \otimes \mathbf{I}_{N_{\mathrm{us}}}$ in the LoS path, and the vectorization form of the input-output relationship in FDA-MIMO-OFDM system becomes
\begin{equation}
    \mathbf{y}_{k,\ell} =  \sum_{p=0}^{P} \tilde{h}_{k,p}\mathbf{b}(\phi_{k,p}) \mathbf{a}^{T}(R_{k,p},\theta_{k,p}) \mathbf{x}_{\ell}.
        \label{Eq3B.2}
\end{equation}
which is consistent with the traditional MIMO-OFDM model under multipath channels.

Nevertheless, the slighter frequency offsets will broaden the beamwidth in the distance domain, thereby reducing the resolution capability of the BS for different UE locations. Adopting symmetric frequency offsets can enhance the distance resolution while maintaining the same degree of ICIs. Let $\rho_{\text{max}}=\max \rho_{m}$ represent the ratio of the maximum frequency offset to the subcarrier spacing. Then, the basic frequency increment cell is $\delta f = 2\rho_{\text{max}} \Delta f/(M-1)$, and the frequency offsets are generated by randomly shuffling the sequence $\left[-(M-1)/2,-(M-1)/2+1,\cdots,(M-1)/2\right]\delta f$. We compare the received symbols on each subcarrier with and without the addition of various frequency offsets in Fig. \ref{Fig.5}. When $\rho_{\text{max}}=1/50$, the two signals are nearly indistinguishable, indicating that approximate orthogonal transmission can be achieved. As $\rho_{\text{max}}=1/50$ increases to 1/20, there is a slight deviation in the received symbols on the part of subcarriers. Significant deviation occurs when $\rho_{\text{max}}=1/10$. At this point, it is necessary to eliminate ICIs to ensure correct system transmission. Taking the subcarrier spacings of 15 kHz and 240 kHz as examples, the beamwidth in the distance domain will reach 200 km and 12.5 km. Such wide beams may only have the possibility of application in satellite communications and it is far beyond the maximum propagation distance designed in Section \ref{Section2A}. Therefore, slight frequency offsets can not meet the design requirements of OFDM waveform, and can only consider the compatibility with other modulation systems, which is beyond the scope of this paper.

\subsection{Pre-equalization}
From \eqref{Eq2C.2}, it can be seen that the precoding matrix $\mathbf{W}$ in an FDA-based system serves a dual purpose:  it is responsible for eliminating IUIs while also maintaining the orthogonality of the subcarriers. To achieve these objectives, we can independently design a pre-equalization matrix followed by traditional precoding at the BS to generate an ICI-free channel. The pre-equalization matrix $\mathbf{P}$ is designed to satisfy the condition $\bar{\mathbf{H}}^{\text{ideal}}= \bar{\mathbf{H}}\mathbf{P}$. Considering the case where the number of antennas $M$ is greater than the number of users $K$, the space-frequency domain
channel matrix $\bar{\mathbf{H}}$ possesses a right inverse. This allows us to derive the pre-equalization matrix as $\mathbf{P}=\bar{\mathbf{H}}^{H} \left(\bar{\mathbf{H}}\bar{\mathbf{H}}^{H}\right)^{-1}\bar{\mathbf{H}}^{\text{ideal}}$. Therefore, the pre-equalization aligns the effective channel seen by each subcarrier with the interference-free scenario, and the performance of FLDMA can be studied with the ideal channel matrix. 

Although it has been established that steering vectors with randomly permuted frequency offsets achieve asymptotic orthogonality across the distance-angle plane, this does not lead to increased DoFs relative to MIMO systems. Given that an array consisting of $M$ elements can only support up to $M$ mutually orthogonal vectors, inter-user interferences persist among UEs positioned at varying angles and distances when $M$ remains finite. Furthermore, the additional spectral overhead and increased noise power exacerbate the performance degradation. Therefore, the potential of FLDMA to enhance spectral efficiency in MIMO systems needs more detailed examination and discussion.

\section{Performance Analysis of FLDMA}

Since the received steering vectors of FLDMA and SDMA are identical, without loss of generality, we consider single-antenna UE for convenience of performance comparison. After adopting the ideal channel matrix, the dimension corresponding to the subcarrier in the MU-MIMO model described in \eqref{Eq2C.2} can be discarded. In this way, the channel matrix for the $k$-th UE 
will degenerate to a $K$-dimensional vector, which is given by $\mathbf{h}^{T}_{k}=\sum_{p=0}^{P} \tilde{h}_{k,p} \mathbf{a}^{T}(R_{k,p},\theta_{k,p})$, and the corresponding MU-MIMO channel model will become to $\bar{\mathbf{H}}=\left[\mathbf{h}_{0},\mathbf{h}_{1},\cdots,\mathbf{h}_{K-1}\right]^{T}\in\mathbb{C}^{K\times M}$. It follows that the transmit symbols degenerate to $\mathbf{s}=\left[s_{0},s_{1},\cdots,s_{K-1}\right]\in\mathbb{C}^{K\times 1}$, and the received symbols similarly degenerate to $\mathbf{y}=\left[y_{0},y_{1},\cdots,y_{K-1}\right]\in\mathbb{C}^{K\times 1}$. Then, ergodic spectrum efficiency in the downlink could be expressed as
\begin{equation}
    \begin{aligned}
       &R^{\text{FLDMA}} = \frac{N_\mathrm{us}}{N_\mathrm{us}+\rho_{\text{max}}}\sum_{k} \mathbf{E}_{\bar{\mathbf{H}}}\left[R_{k}^{\text{FLDMA}}\right]  \\
       &= \frac{N_\mathrm{us}}{N_\mathrm{us}+\rho_{\text{max}}} \mathbf{E}_{\bar{\mathbf{H}}}\left[\sum_{k} \log_{2}\left(1+\dfrac{\left|a_{k}\mathbf{h}^{T}_{k}\mathbf{w}_{k}\right|^{2}}{\sum_{i\neq k} \left|a_{i}\mathbf{h}^{T}_{k}\mathbf{w}_{i}\right|^{2} +\sigma^{2}_{n}}\right)\right].
    \end{aligned}
    \label{Eq4.1}
\end{equation}
MMSE precoding is a commonly employed strategy in downlink multi-user MIMO transmissions, where the transmit precoding vector for each $k$-th UE is determined by
\begin{equation}
    \mathbf{w}_{k} = \frac{1}{\beta_{\text{MMSE}}}\left(\bar{\mathbf{H}}^{H} \bar{\mathbf{H}}+ \dfrac{K\sigma_{n}^{2}}{P}\right)^{-1}\mathbf{h}_{k},
    \label{Eq4.2}
\end{equation}
where $\beta_{\text{MMSE}}$ is the power normalization factor that is set to satisfy $\left\|\mathbf{w}_{k}\right\|=1$. With equal power allocation, the resulting SINR value can be given by \cite{6200372}
\begin{equation}
    \text{SINR}_{k} = \dfrac{\gamma}{\left[\left(\bar{\mathbf{H}}\bar{\mathbf{H}}^{H}+\dfrac{1}{\gamma}\mathbf{I}_{K}\right)^{-1}\right]_{k,k}}-1,
    \label{Eq4.3}
\end{equation}
where $\gamma = \sum_{k}a^{2}_{k}/(K\sigma_{n}^{2})$ is the SNR of each UE. 

Suppose that UEs are uniformly distributed within a sector-shaped region defined by distances from 0 to $R_{\text{max}}$ and angles from $-\theta_{\text{max}}$ to $\theta_{\text{max}}$, where $R_{\text{max}}$ is less than one period of the distance domain and also less than the maximum coverage range permitted by the designed system. In \eqref{Eq4.3}, calculating the SINR for each UE involves inverting matrix $\bar{\mathbf{H}}\bar{\mathbf{H}}^{H}+ 1/\gamma\mathbf{I}_{K}$. We commence by investigating the spectral efficiency by considering the two UEs and each UE operates on a single-path channel for simplicity. Then, the ergodic spectral efficiency can be formulated by the following lemma.
\begin{lemma}
With MMSE precoding and equal power allocation, the ergodic spectrum efficiency of two UEs in a single-path channel can be approximately upper-bounded by

\begin{equation}
    R^{\text{FLDMA, two UEs}} \leq  \frac{2N_\mathrm{us}}{N_\mathrm{us}+\rho_{\text{max}}}\log_{2} \frac{(\gamma+1)^{2}-\gamma^{2}/(M-1)}{1+\gamma}.
        \label{Eq4.4}
\end{equation}

\end{lemma}
\begin{myproof}
Considering two UEs located at $(R_{1},\theta_{1})$ and $(R_{2}, \theta_{2})$, we have
\begin{equation}
\bar{\mathbf{H}}\bar{\mathbf{H}}^{H}+\frac{1}{\gamma}\mathbf{I}_{K} =
\begin{bmatrix}
1+\frac{1}{\gamma} & \eta_{1,2} \\
\eta_{2,1} & 1+\frac{1}{\gamma}
\end{bmatrix}.
\label{Eq4.5}
\end{equation}
Then, the diagonal entries of the inversed matrix are both equal to $(1+1/\gamma)/\left((1+1/\gamma)^2-\left| \eta_{1,2} \right|^{2}\right)-1$, and the spectral efficiency of the two UEs is given by
\begin{equation}
\begin{aligned}
       &R^{\text{FLDMA, two UEs}} \\
       &=  \frac{2N_\mathrm{us}}{N_\mathrm{us}+\rho_{\text{max}}}\mathbf{E}_{\bar{\mathbf{H}}}\left[\log_{2} \frac{\gamma\left((1+1/\gamma)^2-\left| \eta_{1,2} \right|^{2}\right)}{1+1/\gamma}\right] \\
       &\overset{a}{\leq} \frac{2N_\mathrm{us}}{N_\mathrm{us}+\rho_{\text{max}}}\log_{2} \frac{\gamma\left((1+1/\gamma)^2-\mathbf{E}_{\bar{\mathbf{H}}}\left[\left| \eta_{1,2} \right|^{2}\right]\right)}{1+1/\gamma},
        \label{Eq4.6}
\end{aligned}
\end{equation}
where (a) follows from Jensen's inequality, and $\mathbf{E}_{\bar{\mathbf{H}}}\left[\left| \eta_{1,2} \right|^{2}\right]$ can be calculated as
\begin{equation}
    \begin{aligned}
            \mathbf{E}_{\bar{\mathbf{H}}}\left[\left| \eta_{1,2} \right|^{2}\right] &=  \int_{0}^{R_{\text{max}}} \int_{\theta_{\text{max}}}^{\theta_{\text{max}}}  p(R_1 - R_2,\sin{\theta_1} - \sin{\theta_2}) \\
             & \quad \cdot \left| \eta_{1,2} \right|^{2} d (R_{1}-R_{2}) d(\sin{\theta_1} - \sin{\theta_2})) \\
            &\overset{a}{\approx} \dfrac{1}{M-1},
    \end{aligned}
        \label{Eq4.7}
\end{equation}
where the approximation (a) is obtained by $\left|\eta_{i,j}\right| \approx 1/(M-1)$ for most paired $(R_{1},\theta_{1})$ and $(R_{2}, \theta_{2})$, given the extremely low probability of two UEs being located in the same direction or at the same distance. Substituting \eqref{Eq4.7} into \eqref{Eq4.6}, we can obtain \eqref{Eq4.4}, which completes the proof.
\end{myproof}

Meanwhile, we are interested in whether FLDMA enhances system capacity compared to SDMA. Observing \eqref{Eq4.6}, the spectral efficiency is equivalent to that of SDMA when $\delta f =0$. Since the number of subcarriers $N_{\mathrm{us}}$ does not affect the calculation of SINR, the frequency offsets overhead of FLDMA with infinite bandwidth can be negligible. Along with the above analysis, the gap between the two systems can be compared directly by the beam correlation.
\begin{corollary}{\label{corollary3}}
In a single-path downlink channel with infinite bandwidth, the FLDMA with 2 UEs outperforms SDMA in spectral efficiency when ${\rm Sa}^2_{M} \left((\sin{\theta_1}-\sin{\theta_2})/2) \right)>1/M$ is satisfied.
\end{corollary}
\begin{myproof}
    The beam correlation of SDMA can be directly determined by $\left|\eta^{\text{SDMA}}_{1,2}\right|^2={\rm Sa}^2_{M} \left(p_{1,2} \right)$, so we can make the following comparison
    \begin{equation}
        \begin{aligned}
            &\left|\eta^{\text{FLDMA}}_{1,2}\right|^2 - \left|\eta^{\text{SDMA}}_{1,2}\right|^2 \\
            &= \dfrac{1}{M} +\dfrac{M}{M-1} \left({\rm Sa}_{M}^{2}\left(p_{1,2}\right)-\frac{1}{M}\right)\left({\rm Sa}_{M}^{2}\left(q_{1,2}\right) -\frac{1}{M}\right) \\
            & \quad - {\rm Sa}^2_{M} \left(p_{1,2} \right) \\
            & = \left(\dfrac{M{\rm Sa}^2_{M} \left(p_{1,2} \right)-1}{M-1}\right)\left({\rm Sa}^2_{M} \left(q_{1,2} \right)-1\right).
        \end{aligned}
    \end{equation}
    The performance of FLDMA exceeds that of SDMA when $ \left|\eta^{\text{FLDMA}}_{1,2}\right|^2 < \left|\eta^{\text{SDMA}}_{1,2}\right|^2$. Since ${\rm Sa}^2_{M} \left(q_{1,2} \right)-1 \leq 0$, it means ${\rm Sa}^2_{M} \left((\sin{\theta_1}-\sin{\theta_2})/2) \right)>1/M$ needs to be satisfied, which completes the proof.
\end{myproof}
\textbf{Corollary \ref{corollary3}} illustrates that in a downlink scenario with only two UEs, the performance of the FLDMA system surpasses that of SDMA only when the UEs are located within a certain threshold of $\theta_{\text{max}}$.

Subsequently, we further analyze the spectrum efficiency of multiple UEs. 
At this point, $\left|\eta_{i,j}\right|^2$ cannot be briefly approximated as $1/M$. To our best knowledge, it is still intractable to derive an analytical solution for the diagonal elements of such inversed Hermitian matrices. The statistical properties of $\left[\left(\bar{\mathbf{H}}\bar{\mathbf{H}}^{H}+1/\gamma\mathbf{I}_{K}\right)^{-1}\right]_{k,k}$ in the case of $\bar{\mathbf{H}}$ being i.i.d. Gaussian matrix have been analyzed in \cite{7166320}. The maximum spectrum efficiency of multiple UEs can be approximatively obtained through the following lemma.
\begin{lemma}{\label{lemma3}}
With the high SNR assumption, the spectral efficiency of FLDMA can be approximately upper-bounded by
\begin{equation}
    R^{\text{FLDMA}} \leq \frac{KN_\mathrm{us}}{N_\mathrm{us}+\rho_{\text{max}}} \log_{2}\dfrac{\gamma(M-K+1)}{M}.
\end{equation}
\end{lemma}
\begin{myproof}
    Using Jensen's inequality, we can obtain
    \begin{equation}
    \begin{aligned}
        R^{\text{FLDMA}} \overset{\text{High SNR}}{\approx} &\frac{N_\mathrm{us}}{N_\mathrm{us}+\rho_{\text{max}}}\sum_{k}\mathbf{E}_{\bar{\mathbf{H}}}\left[\log_{2}\dfrac{\gamma}{[(\bar{\mathbf{H}}\bar{\mathbf{H}}^{H})^{-1}]_{k,k}} \right]\\
        \leq &\frac{N_\mathrm{us}}{N_\mathrm{us}+\rho_{\text{max}}}\sum_{k}\log_{2}\left(\gamma\mathbf{E}_{\bar{\mathbf{H}}}\left[\dfrac{1}{[(\bar{\mathbf{H}}\bar{\mathbf{H}}^{H})^{-1}]_{k,k}}\right]\right).
        \end{aligned}
    \end{equation}
As indicated in \cite{wong2008array}, if the channel correlation between different UEs is 0 and the independence between array elements holds in an MU-MIMO channel, it follows that $\mathbf{E}_{\bar{\mathbf{H}}}\left[1/[(\bar{\mathbf{H}}\bar{\mathbf{H}}^{H})^{-1}]_{k,k}\right]=(M-K+1)/K$. Obviously, the UEs are independently distributed, thus satisfying the condition of correlation. As for $\mathbf{E}_{\bar{\mathbf{H}}} \left[\mathbf{h}^H_{i} \mathbf{h}_{i}\right]$, the diagonal elements are absolutely $1/M$, and the elements in the $n$-th row and $m$-th column can be expressed by
\begin{equation}
    \begin{aligned}
        \mathbf{E}_{\bar{\mathbf{H}}}\left[\mathbf{h}^H_{i}\mathbf{h}_{i}\right]_{n,m} = &\mathbf{E}_{\bar{\mathbf{H}}}\left[\sum_{p}\sum_{l} h^{*}_{l}h_{p}e^{j\pi (m\sin\theta_p-n\sin\theta_l)} \right. \\
        & \left. e^{j2\pi\delta f\frac{R_{l}z_n-R_{p}z_m}{c}}\right].
    \end{aligned}
        \label{Eq4.8}
\end{equation}
Since all variables in \eqref{Eq4.8} are independent and $R_{p}$ follows uniform distribution, we can immediately obtain $\mathbf{E}_{\bar{\mathbf{H}}} \left[\mathbf{h}^H_{i} \mathbf{h}_{i}\right]_{n,m} =0$, which completes the proof. 
\end{myproof}
Additionally, it is apparent that \eqref{Eq4.8} still holds under a single-path channel. This finding highlights that FLDMA's system performance is inherently independent of multipath channel effects, which facilitates full MIMO channel degrees of freedom for multiple access. Such a characteristic is particularly advantageous in the millimeter-wave and higher frequency domains, where multipath effects can severely limit the capabilities of SDMA systems.

\section{Simulation Results}
In this section, we evaluate the performance of the proposed FLDMA scheme through numerical simulations. To satisfy the approximation in \eqref{Eq2A.6}, the carrier frequency is set to 30 GHz. The array is configured with a half-wavelength spacing, resulting in a distance of $d=0.5$cm. The subcarrier spacing is set to be 15 kHz with $N_\mathrm{us}=512$, and the CP length is 1/4 OFDM symbol period, which is about $16.67\mathrm{\mu s}$. As outlined in Section \ref{Section2A}, the coverage area of the designed system should ensure that the maximum channel propagation delay remains within the confines of the CP length. Assuming the maximum multipath delay spread is $6.67\mathrm{\mu s}$, it yields a maximum LoS coverage distance of 3 km for the BS. Therefore, the UEs are set to be uniformly distributed within the sector area defined by $\theta \in [-\theta_{\text{max}},\theta_{\text{max}}]$ and $R \in [0, 3 \mathrm{km}]$. The BS is equipped with a ULA of 128 elements, while the UEs are equipped with a single antenna in simulations. All simulations come from 10,000 Monte Carlo samples.

\subsection{Two UEs under the Single-path Channel}
\begin{figure}[tbp]
    \centering
    \subfigure[$\theta_{\text{max}}=5^\circ$]{
        \begin{minipage}[t]{1\linewidth}
            \centering
            \includegraphics[width=3.5in]{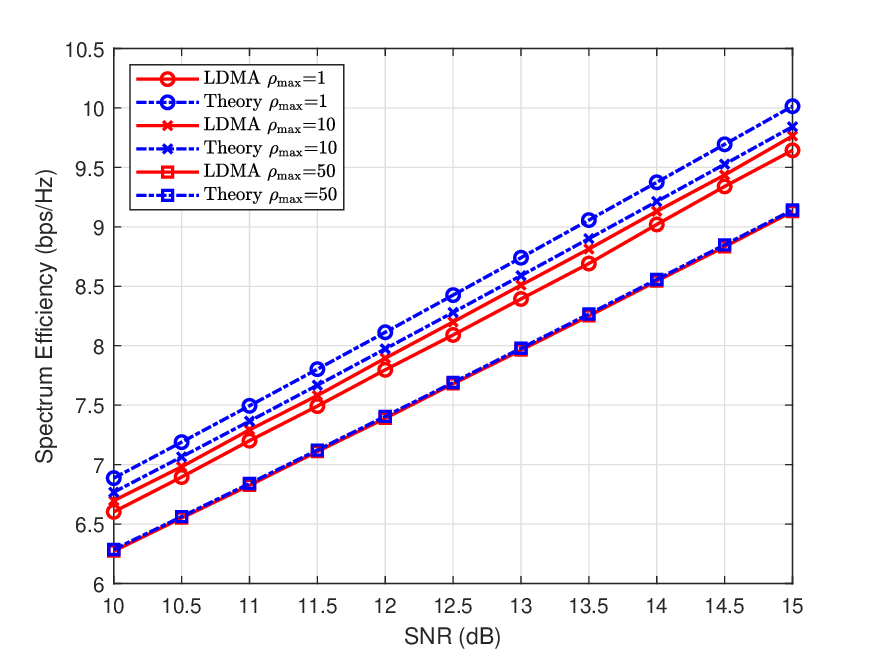}
            \label{Fig.6a}
        \end{minipage}%
    }%
    
    \subfigure[$\theta_{\text{max}}=20^\circ$]{
        \begin{minipage}[t]{1\linewidth}
            \centering
            \includegraphics[width=3.5in]{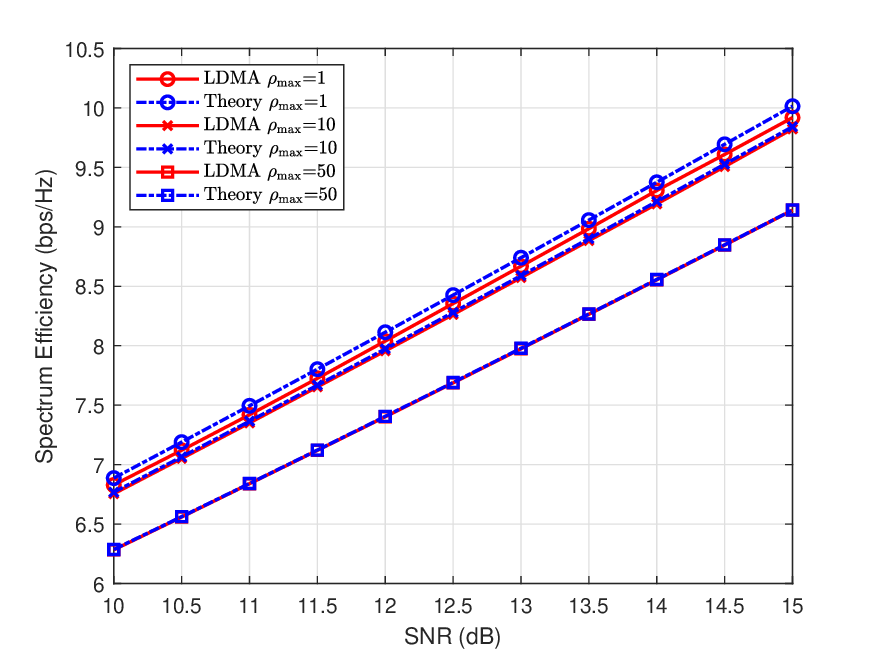}
            \label{Fig.6b}
        \end{minipage}%
        }
    \caption{Spectral efficiency achieved in a single-path channel in FLDMA system consisting of two UEs.}
    \label{Fig.6}
\end{figure}
First, the curves of spectral efficiency versus SNRs are plotted in Fig. \ref{Fig.6}. We establish two simulations where the coverage angles of the BS are $[-5^\circ, 5^\circ]$ and $[-20^\circ, 20^\circ]$. In each simulation, we consider three frequency offset ratios $\rho _{\text{max}}=1,10,50$, corresponding to distance-domain beam resolutions of 20 km, 2 km, and 400 m, respectively. The red represents the simulation results, and the blue lines are evaluated by \eqref{Eq4.4}. We can notice that as $\theta_{\text{max}}$ or $\rho_{\text{max}}$ increases, the driver upper bound gradually becomes tighter. With a larger $\theta_{\text{max}}$, the probability of different UEs being located in the same direction or at the same distance decreases, thereby rendering the approximation $\left|\eta_{i,j}\right| \approx 1/(M-1)$ more reasonable. Likewise, the beam resolution in the distance domain will improve as $\rho_{\text{max}}$ increases, so the BS acquires a higher precision to distinguish between UEs that are close to each other. Furthermore, the simulated spectral efficiency for $\rho_{\text{max}}=10$ at $\theta_{\text{max}}=5^\circ$ is slightly higher than that for $\rho_{\text{max}}=1$, whereas the opposite is true at $\theta_{\text{max}}=20^\circ$. This observation implies that there exists an ideal value of $\rho_{\text{max}}$ that maximizes system performance within the constraints of a specified $\theta_{\text{max}}$, pointing to the demand for parameter tuning for optimal transmission. Moreover, the spectral efficiency for $\rho_{\text{max}}=50$ is the lowest, indicating that higher distance resolution does not necessarily provide better system performance. The fundamental reason is that the SINR gains afforded by a larger $\rho_{\text{max}}$ fail to offset the spectral losses.

\begin{figure}[tbp]
    \includegraphics[width=3.5in]{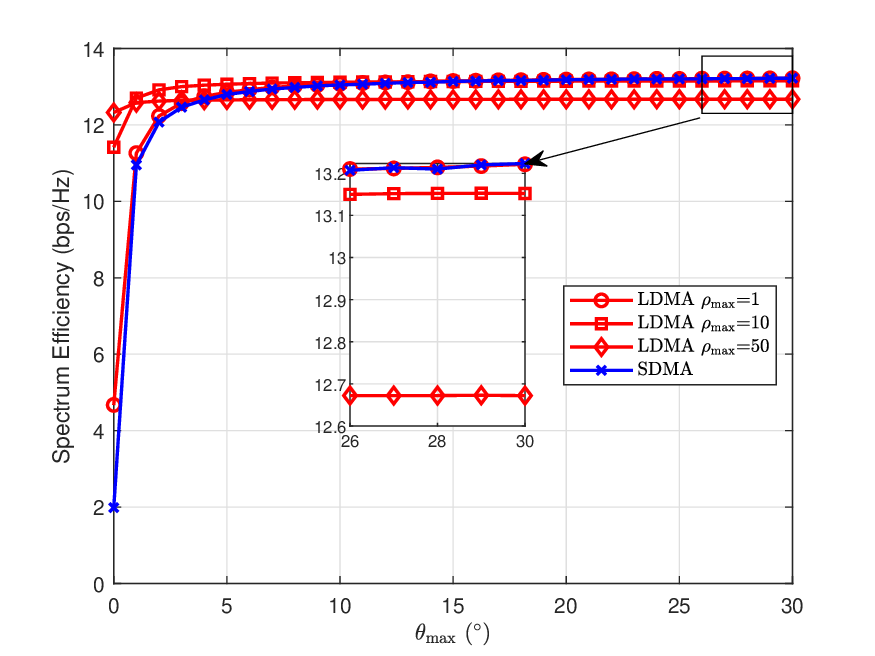}
    \centering
    \caption{Comparison of the proposed FLDMA with various $rho_{\text{max}}$ and SDMA for different $\theta_{\text{max}}$ in a system with two UEs.}
    \label{Fig.7}
\end{figure}
To further explore the performance of FLDMA, the spectrum efficiency with increasing $\theta_{\text{max}}$ under various values for $\rho_{\text{max}}$ is shown in Fig. \ref{Fig.7}, and makes a comparison with SDMA. SNR is set to be 20 dB. When $\theta_{\text{max}}\leq 3^{\circ}$, particularly at $\theta_{\text{max}}=0^{\circ}$ where UEs are aligned in a straight line, the performance of the FLDMA system with $\rho_{\text{max}}=10$ and $\rho_{\text{max}}=50$ substantially surpasses that of SDMA. This demonstrates FLDMA's advanced capability to effectively reduce IUIs under such location configurations. However, once $\theta_{\text{max}}\geq 15^{\circ}$, the spectral efficiency of SDMA will exceed that of the above two systems, which is consistent with \textbf{Corollary \ref{corollary3}}. One aspect is that while interference in SDMA diminishes as the angle between UEs increases, the interference in our proposed FLDMA remains relatively stable even with an increase in the Euclidean distance between UEs, unless they coincide at the same angle or distance. On the other hand, FLDMA has extra spectral overhead, inherently placing it at a disadvantage in scenarios with fewer users.
Given that UEs are situated within the confines of a single beamwidth of distance domain, it becomes evident that the performance of the FLDMA system with $\rho_{\text{max}}=1$ closely matches that of SDMA, exhibiting a more significant performance improvement only at $\theta_{\text{max}}=0^{\circ}$. If UEs are permitted to be distributed across a wider distance, employing $\rho_{\text{max}}=1$ can significantly increase the BS's probability of successfully distinguishing between two UEs' locations, leading to enhanced spectral efficiency. From this, it follows that under the conditions of a specified number of subcarriers and UE distribution range, there theoretically exists an optimal $\rho_{\text{max}}$.

\subsection{Multiple UEs under the Multipath channel}

\begin{figure}[tbp]
    \includegraphics[width=3.5in]{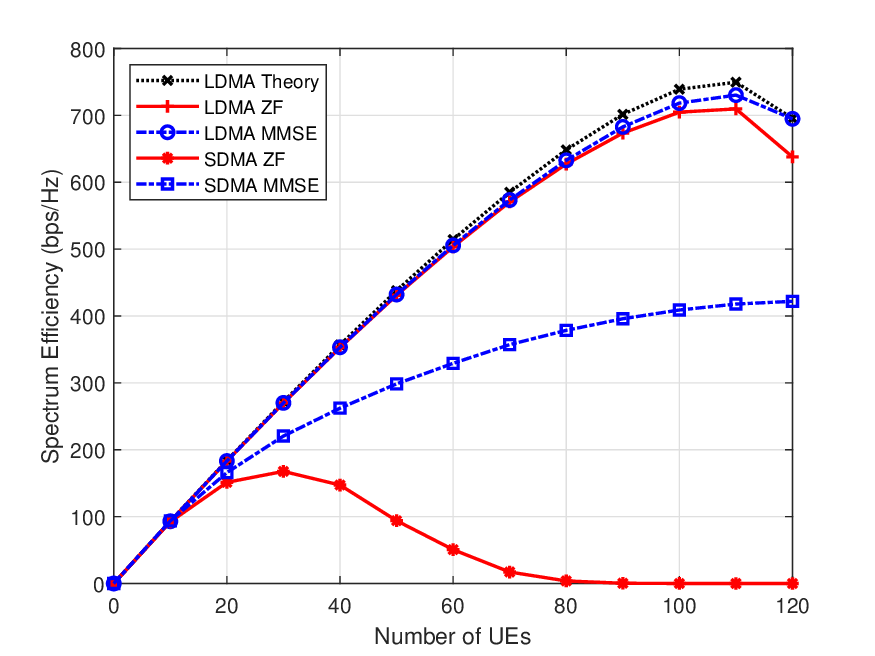}
    \centering
    \caption{Comparison of the proposed FLDMA and SDMA for different numbers of UEs under single-path channel.}
    \label{Fig.8}
\end{figure}

\begin{figure}[tbp]
    \includegraphics[width=3.5in]{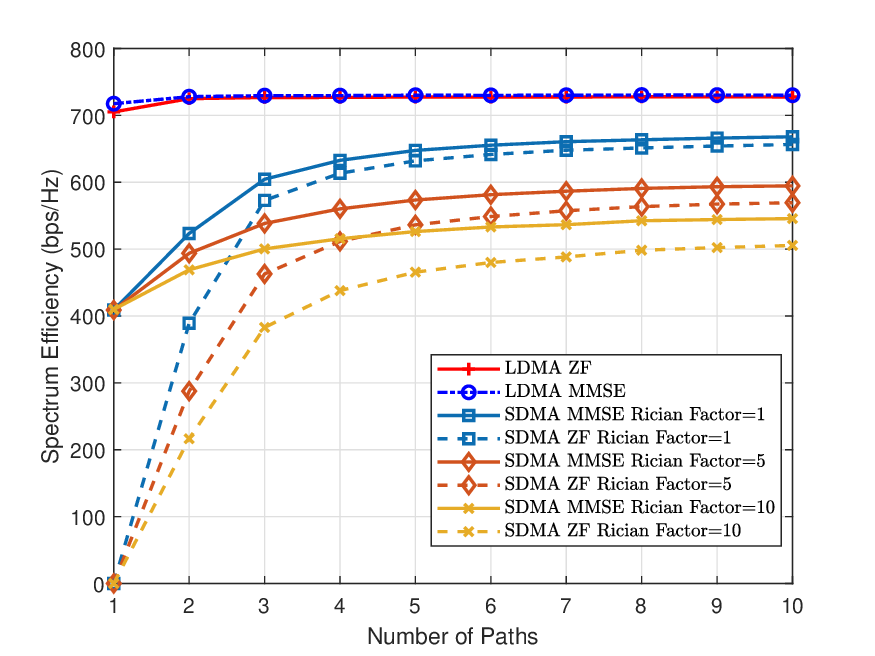}
    \centering
    \caption{Comparison of the proposed FLDMA and SDMA for different numbers of path channels.}
    \label{Fig.9}
\end{figure}
\begin{figure}[tbp]
    \includegraphics[width=3.5in]{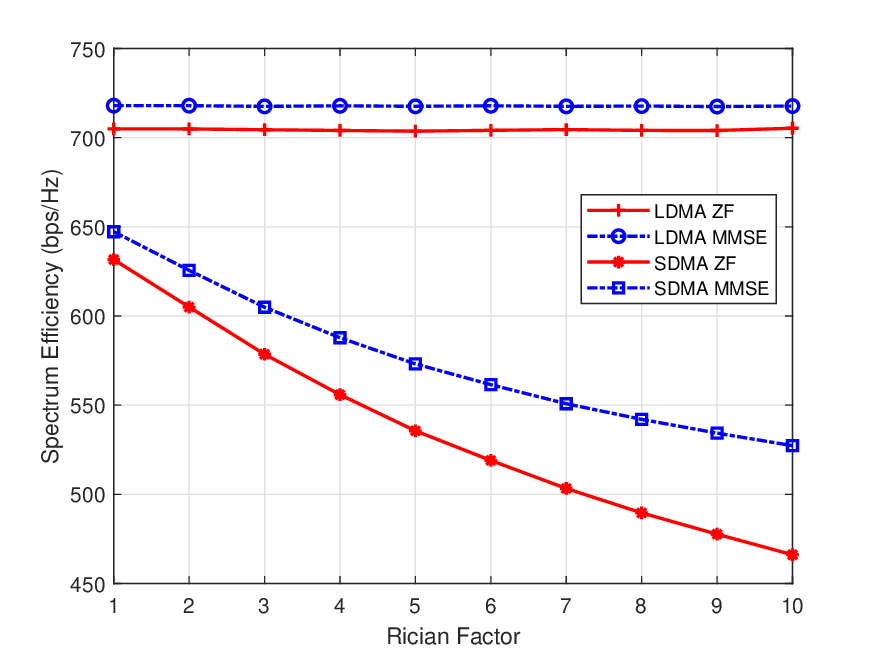}
    \centering
    \caption{Comparison of the proposed FLDMA and SDMA for different Rician factors.}
    \label{Fig.10}
\end{figure}

To validate the performance improvement of the proposed FLDMA over SDMA in multiple UEs scenarios, we first depict the spectral efficiency of systems using MMSE and ZF precoding under a single-path channel Fig. \ref{Fig.7}. The SNR is configured at 30 dB, and $\rho_{\text{max}}=30$. UEs are distributed within a sector spanning from $-60^{\circ}$ to $60^{\circ}$ degrees. The theoretical bound derived from \textbf{Lemma \ref{lemma3}} is plotted with a black dashed line. We observe that FLDMA's spectral efficiency begins to surpass that of SDMA once the number of UEs exceeds 10. For SDMA with ZF precoding, spectral efficiency starts to decline when the number of UEs reaches 30 and tends toward zero upon reaching 80. Even with MMSE precoding, SDMA's spectral efficiency only increases slowly as the number of UEs grows. However, the spectral efficiency of FLDMA systems utilizing ZF and MMSE precoding increases rapidly and aligns closely with theoretical values until the number of UEs reaches 100. There is a noticeable divergence from the theoretical bound only when the UE count exceeds 80. Therefore, the proposed FLDMA can provide about $75\%$ performance gain compared with SDMA at SNR=30 dB. The substantial performance improvement can be attributed to FLDMA's ability to effectively distinguish UEs at different distances. However, the fundamental reason lies in the newly designed steering vectors, which exhibit more uniform beam correlation across each pair of UEs compared to traditional PA steering vectors. In the SDMA system, the steering vectors exhibit high correlation when the angles between two UEs are close, resulting in some very small eigenvalues of $\bar{\mathbf{H}}\bar{\mathbf{H}}^{H}$. This is reflected in the inversed matrix where some diagonal elements become significantly large. During the Frobenius norm normalization process, a considerable amount of transmit power is allocated to these larger diagonal elements, leading to substantial disparities in the SINR among UEs. Conversely, the diagonal elements of the inversed matrix are relatively uniform in FLDMA, thereby achieving higher spectral efficiency.

\begin{figure}[tbp]
    \centering
    \subfigure[40 UEs]{
        \begin{minipage}[t]{1\linewidth}
            \centering
            \includegraphics[width=3.5in]{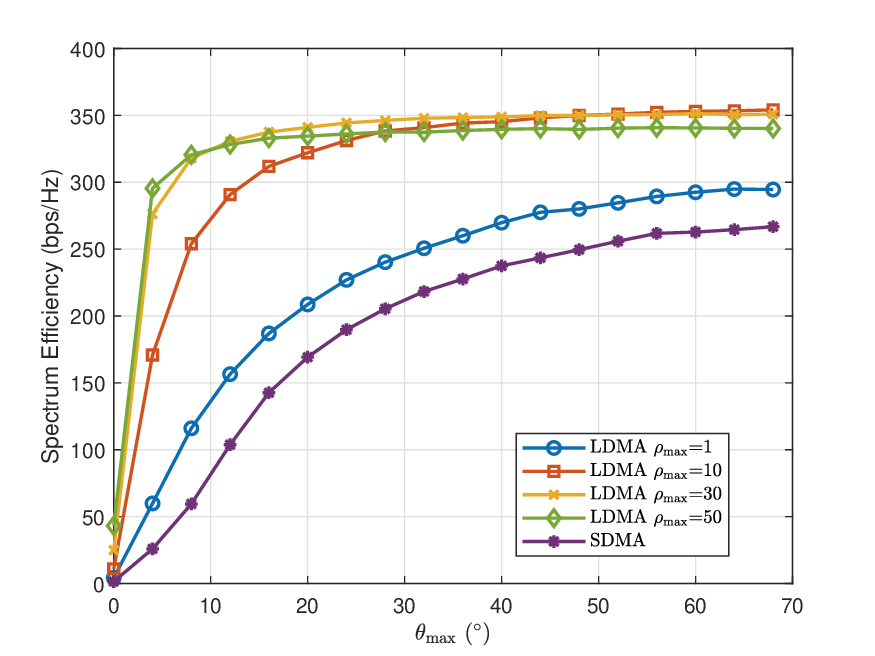}
            \label{Fig.11a}
        \end{minipage}%
    }%
    
    \subfigure[100 UEs]{
        \begin{minipage}[t]{1\linewidth}
            \centering
            \includegraphics[width=3.5in]{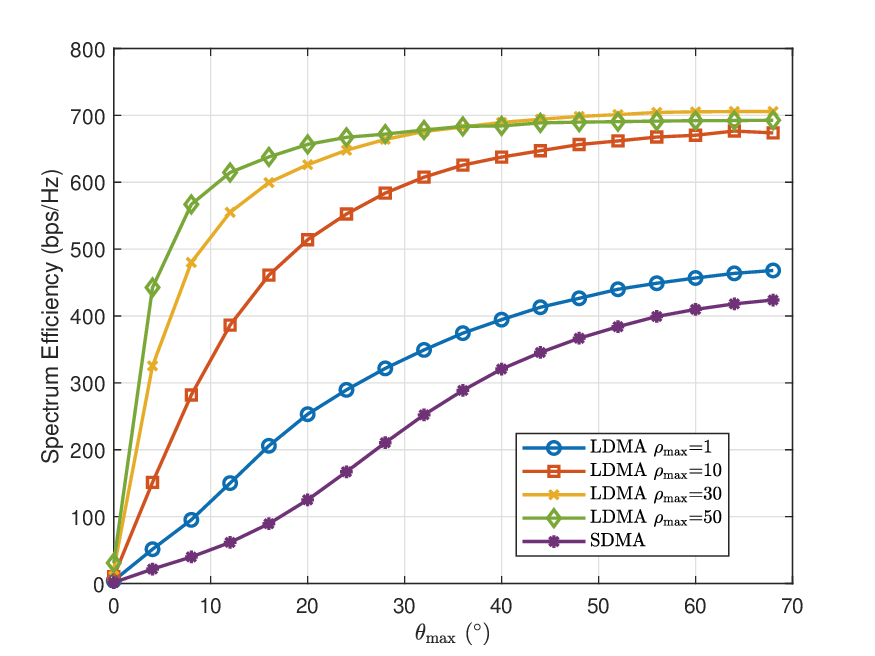}
            \label{Fig.11b}
        \end{minipage}%
        }
    \caption{Spectral efficiency versus $\theta_{\text{max}}$ for proposed FLDMA with various $\rho_{\text{max}}$ and SDMA.}
    \label{Fig.11}
\end{figure}

In Fig. \ref{Fig.9}, we compare the performance of FLDMA and SDMA among 100 UEs across different numbers of paths $P$. With the increase of $P$, the spectral efficiency of SDMA gradually increases, while that of FLDMA remains constant. Additionally, it becomes apparent that the performance of SDMA progressively deteriorates as the Rician factor increases. This is better illustrated in Fig. \ref{Fig.10}, where it is evident that FLDMA's performance is mostly unaffected by changes in the Rician factor. The underlying reasons for these phenomena are that the channel matrix elements of SDMA gradually conform to i.i.d. Gaussian due to the Central Limit Theorem in rich scattering environments. As a result, SDMA's performance continues to improve with the increasing $P$. Conversely, FLDMA's channel matrix meets the independence conditions even in a single-path scenario, which is why the performance of FLDMA tends towards the upper bound as shown in Fig. \ref{Fig.8}. This indicates that FLDMA's system performance does not depend on multipath effects, thereby enabling ideal spectral efficiency in millimeter waves.

\begin{figure}[tbp]
    \centering
    \subfigure[40 UEs]{
        \begin{minipage}[t]{1\linewidth}
            \centering
            \includegraphics[width=3.5in]{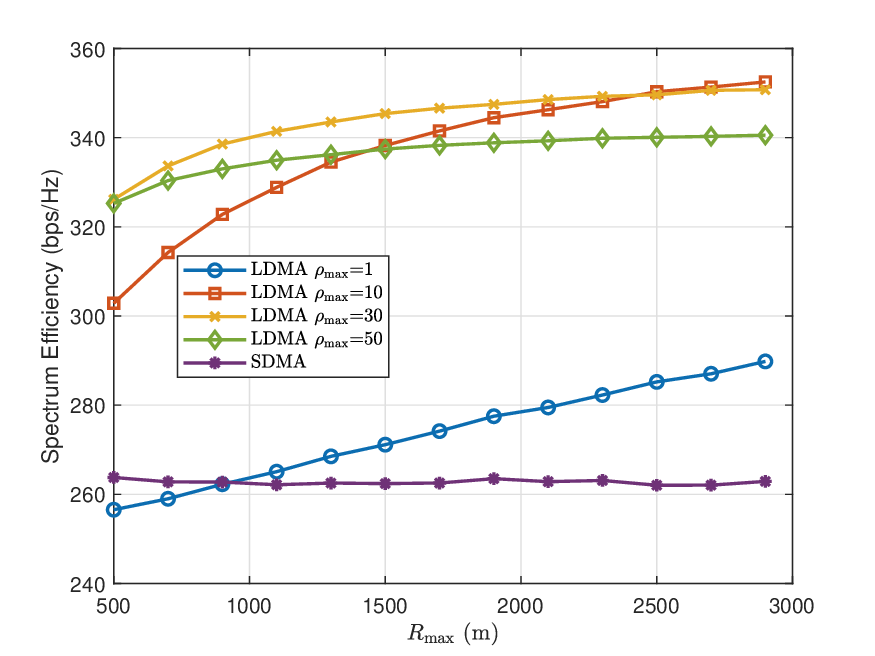}
            \label{Fig.12a}
        \end{minipage}%
    }%
    
    \subfigure[100 UE]{
        \begin{minipage}[t]{1\linewidth}
            \centering
            \includegraphics[width=3.5in]{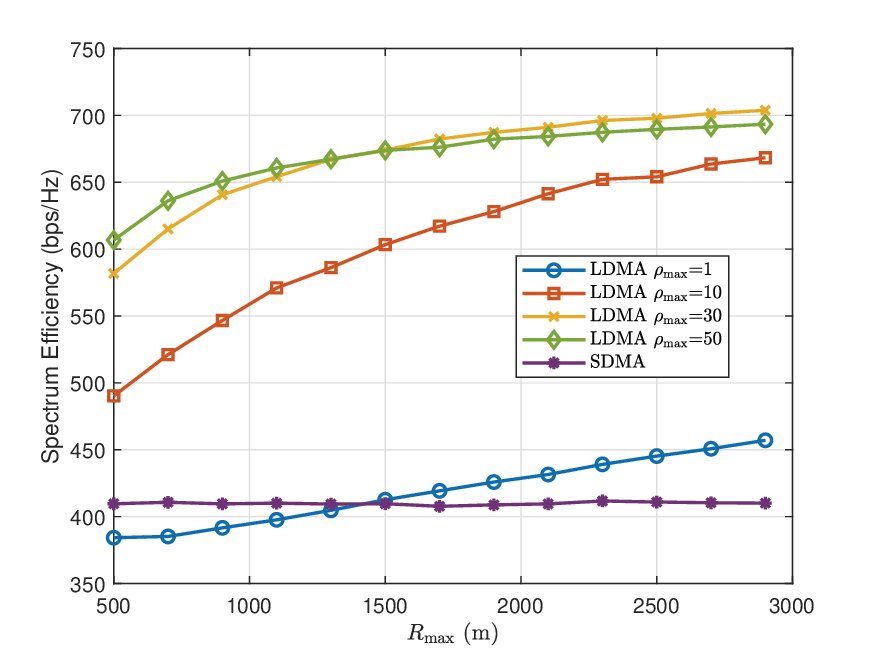}
            \label{Fig.12b}
        \end{minipage}%
        }
    \caption{Spectral efficiency versus $R_{\text{max}}$ for proposed FLDMA with various $\rho_{\text{max}}$ and SDMA.}
    \label{Fig.12}
\end{figure}

Finally, we further explored the variation of spectral efficiency by adjusting $\theta_{\text{max}}$ and $R_{\text{max}}$. Fig. \ref{Fig.11} elucidates that there consistently emerges an optimal value for $\rho_{\text{max}}$ that is capable of optimizing spectral efficiency across a spectrum of angle settings and varying numbers of UEs. A similar phenomenon is also evident when altering $R_{\text{max}}$, as demonstrated in Fig. \ref{Fig.12}. As previously mentioned, the performance enhancements of FLDMA stem from its ability to distribute interference more evenly. Elevating $\rho_{\text{max}}$ serves to narrow the mainlobe width in the two-dimensional plane, reducing the probability of UEs sharing the same beam and thus boosting spectral efficiency. However, increasing $\rho_{\text{max}}$ also leads to a rise in frequency overhead. It follows that under specific conditions for $\theta_{\text{max}}$, $R_{\text{max}}$ and the number of UEs, there exists an optimal value of $\rho_{\text{max}}$ that can maximize spectral efficiency. Moreover, as shown in Fig. \ref{Fig.11}, it can be observed that the spectral efficiency initially increases sequentially with the rise of $\theta_{\text{max}}$, and then converges beyond a certain threshold. Therefore, FLDMA exhibits several times the spectral efficiency of SDMA in ultra-dense networks with narrow sectors. A consequent idea is to adopt FLDMA after dividing the cellular network into smaller sectors, which can significantly improve the system capacity of the BS.

\section{Conclusion}
In this paper, we propose a far-field FLDMA scheme based on the ability of the FDA to manipulate far-field beams in the distance domain. By using random permutation frequency offsets, the steering vectors also satisfy asymptotic orthogonality in the distance-angle domain. A pre-equalization method is proposed to eliminate ICIs caused by frequency offsets, which can achieve ideal orthogonal transmission in the MU-MIMO system. Simulation results show that the proposed FLDMA scheme has significant spectral efficiency advantages, especially in narrow sector multiple communication scenarios. The independence between antennas in FLDMA makes it highly promising for future millimeter-wave and higher-frequency communication systems. For future research, since the design principles of OFDM limit the maximum coverage of FLDMA, other waveforms compatible with FDA need to be considered for longer-distance communications. Furthermore, the optimal $\rho_{\text{max}}$ for different communication scenarios is another important direction for future research.
\section*{Acknowledgment}

\ifCLASSOPTIONcaptionsoff
  \newpage
\fi





\bibliographystyle{IEEEtran}
\bibliography{Bibliography}

\vfill


\end{document}